\documentclass[twocolumn]{aastex631}

\usepackage{amssymb,amsmath,amsfonts}
\usepackage{graphicx}
\usepackage{xcolor}
\usepackage{epsfig}
\usepackage{longtable}
\usepackage{verbatim}
\usepackage{color}
\usepackage{mdframed}
\usepackage{soul} 
\usepackage{braket}
\usepackage{enumitem}
\usepackage[percent]{overpic}
\usepackage{float}
\usepackage{array,booktabs,tabularx}
\newcolumntype{C}{>{\centering\arraybackslash}X}
\renewcommand{\arraystretch}{1.6}
\usepackage{soul}
\usepackage{placeins}
\include{eps2pdf}

\usepackage[flushmargin,bottom]{footmisc}
\newcommand{ \e}{\mathrm{e}^ }
\newcommand{\be}{\begin{equation}}
\newcommand{\ee}{\end{equation}}
\newcommand{\bea}{\begin{eqnarray}}
\newcommand{\eea}{\end{eqnarray}}

\def\k{{\bf k}}
\def\q{{\bf q}}
\def\x{{\bf x}}
\def\d{{\rm d}}

\shorttitle{Neutrino properties from Line Intensity Mapping}
\shortauthors{Moradinezhad Dizgah \& Keating et al.}

\begin{document} 

\title{Neutrino Properties with Ground-Based Millimeter-Wavelength Line Intensity Mapping} 

\author{Azadeh Moradinezhad Dizgah $^{\ast}$}
\affiliation{D\'epartement de Physique Th\'eorique,
Universit\'e de Gen\`eve, 24 quai Ernest Ansermet, 1211 Gen\`eva 4, Switzerland}

\author{Garrett~K.~Keating $^{\ast \ast}$}
\affiliation{Center for Astrophysics, Harvard \& Smithsonian, 60 Garden Street, Cambridge, MA 02138, USA}

\author{Kirit~S.~Karkare}
\affiliation{Kavli Institute for Cosmological Physics, University of Chicago, 5640 S. Ellis Ave., Chicago, IL 60637, USA}
\affiliation{Fermi National Accelerator Laboratory, MS209, P.O. Box 500, Batavia, IL 60510, USA}

\author{Abigail~Crites}
\affiliation{Department of Physics, Cornell University, 109 Clark Hall, Ithaca, NY 14853, USA }
\affiliation{Department of Physics, California Institute of Technology, 1200 E. California Blvd., Pasadena, CA 91125, USA }

\author{Shouvik~Roy~Choudhury}
\affiliation{Department of Physics, Indian Institute of Technology Bombay, Main Gate Road, Powai, Mumbai 400076, India \\ }

\email{$^{\ast}$ azadeh.moradinezhaddizgah@unige.ch}
\email{$^{\ast \ast}$ garrett.keating@cfa.harvard.edu}

\begin{abstract}
Line intensity mapping (LIM) is emerging as a powerful technique to map the cosmic large-scale structure and to probe cosmology over a wide range of redshifts and spatial scales. We perform Fisher forecasts to determine the optimal design of wide-field ground-based mm-wavelength LIM surveys for constraining properties of neutrinos and light relics. We consider measuring the auto-power spectra of several CO rotational lines (from J=2-1 to J=6-5) and the [CII] fine-structure line in the redshift range of $0.25<z<12$. We study the constraints with and without interloper lines as a source of  noise in our analysis, and for several one- and multi-parameter extensions of $\Lambda$CDM. We show that LIM surveys deployable this decade, in combination with existing CMB (primary) data, could achieve  order of magnitude improvements over Planck constraints on $N_{\rm eff}$ and $M_\nu$. Compared to next-generation CMB and galaxy surveys, a LIM experiment of this scale could achieve bounds that are a factor of $\sim3$ better than those forecasted for surveys such as Euclid (galaxy clustering), and potentially exceed the constraining power of CMB-S4 by a factor of $\sim1.5$ and $\sim3$ for $N_{\rm eff}$ and $M_\nu$, respectively. We show that the forecasted constraints are not substantially affected when enlarging the parameter space, and additionally demonstrate that such a survey could also be used to measure $\Lambda$CDM parameters and the dark energy equation of state exquisitely well. \\
\end{abstract}


\section{Introduction}

Neutrinos are among the most abundant particles in the Universe, and thus affect different epochs in the cosmic history. Cosmological observations are sensitive to the effective number of neutrinos, $N_{\rm eff}$, when they were still relativistic and contributed to the radiation content of the Universe, as well as their total mass, $M_\nu \equiv \Sigma m_\nu$, when they became non-relativistic and contribute to the matter content \citep{Abazajian:2016yjj,Lattanzi:2017ubx,Lesgourgues:2018ncw}. The standard model of particle physics predicts three species of massless neutrinos, corresponding to $N_{\rm eff}^{\rm SM} = 3.046$ \citep{Mangano:2005cc,Grohs:2015tfy,deSalas:2016ztq}. While the current best constraint on $N_{\rm eff}$ from the Cosmic Microwave Background (CMB) \citep{Aghanim:2018eyx} is consistent with this prediction, neutrino flavor oscillation experiments, which have measured non-zero neutrino mass splittings \citep{deSalas:2017kay,Esteban:2018azc}, provide  striking evidence in favor of physics beyond the standard model (BSM) to describe the origin of non-zero neutrino masses.

Next-generation CMB experiments \citep{SimonsObservatory:2018koc,Abazajian:2016yjj} are expected to provide stringent constraints on $N_{\rm eff}$ \citep{Green:2019glg}. Observations of large-scale structure (LSS) will be complementary to CMB, particularly in ameliorating parameter degeneracies (e.g., between $N_{\rm eff}$ and $\Lambda$CDM parameters as well as with the sum of neutrino masses and primordial Helium abundance \citep{Baumann:2017gkg,Sprenger:2018tdb,DePorzio:2020wcz}). Since $N_{\rm eff}$ measures the total energy density in radiation excluding photons, a high-significance detection of an excess light relic abundance, $N_{\rm eff} = N_{\rm eff}^{\rm SM} + \Delta N_{\rm eff}$, offers a discovery space for BSM physics: many extensions to the SM predict extra light relics, for example axions \citep{Baumann:2016wac} and light sterile neutrinos \citep{Abazajian:2012ys,Archidiacono:2014nda}. The contribution of light thermal relics to $N_{\rm eff}$ is determined by their number of spin states and decoupling temperatures. As such, there is a minimum contribution to $N_{\rm eff}$ from light relics that decoupled prior to the QCD phase transition, $\Delta N_{\rm eff} \geq 0.027$ \citep{Brust:2013ova,Chacko:2015noa}, which sets a theoretical target sensitivity for upcoming surveys. 

In contrast to $N_{\rm eff}$, constraining the total mass of neutrinos from CMB primary anisotropies is challenging since neutrinos with sub-eV mass are still relativistic around the last scattering surface, and thus their mass affects the CMB weakly. On the other hand, gravitational lensing of the CMB, which indirectly probes the underlying dark matter distribution, is a sensitive probe of neutrino masses. But several parameter degeneracies---most notably between $M_\nu$, the total matter density and optical depth \citep{Allison:2015qca,Liu:2015txa}, and between $M_\nu$ and the dark energy (DE) equation-of-state (EoS) \citep{Hannestad:2005gj,RoyChoudhury:2019hls}---limit the potential of upcoming CMB surveys \citep{SimonsObservatory:2018koc,Abazajian:2016yjj,Sugai:2020pjw} in constraining $M_\nu$. Biased tracers of LSS provide the most promising window to probe massive neutrinos via their imprints on the expansion history and growth of structure using various statistics, including auto- and cross-correlations between different probes \citep{Dvorkin:2019jgs,Boyle:2017lzt,Schmittfull:2017ffw,Yu:2018tem,Chudaykin:2019ock,Hahn:2019zob,Boyle:2020rxq,Massara:2020pli,Hahn:2020lou,Bayer:2021iyb,Chen:2021vba}. The measurements of mass splitting by neutrino flavor oscillation experiments\footnote{Neutrino oscillation experiments to date \citep{deSalas:2017kay,Esteban:2018azc} have measured two squared-mass differences between the three neutrino species, allowing for two possible mass hierarchies: normal (two light and one heavy) vs. inverted (two heavy and one light). While recent results show a weak preference for the normal ordering \citep{Super-Kamiokande:2017yvm,T2K:2018rhz,NOvA:2019cyt}, future neutrino experiments such as DUNE \citep{DUNE:2020jqi} and Hyper-K \citep{Hyper-Kamiokande:2018ofw} promise an unambiguous determination of the two hierarchies (see \citealt{Patterson:2015xja} for a review of the experimental prospects).}, which set a minimum for the sum of neutrino masses, $M_\nu = 0.059\ {\rm eV}$ in the normal and $M_\nu = 0.10 \ {\rm eV}$ in the inverted mass hierarchies, provide theoretical thresholds for future cosmological measurements of $M_\nu$.

In addition to shape and clustering of galaxies and CMB secondary anisotropies, line intensity mapping (LIM) is emerging as a viable probe of LSS \citep{Kovetz:2017agg}. Measuring spatial fluctuations in the brightness temperature of spectral lines together with their observed frequencies provides a low-resolution, 3-dimensional map of LSS.  LIM experiments can efficiently survey large sky fractions and extended redshift ranges, largely inaccessible to traditional galaxy surveys. The promise of LIM in constraining cosmological parameters is threefold. First, the large comoving volume probed by LIM surveys significantly lowers statistical uncertainty on model parameters. Second, at the higher redshifts uniquely probed by LIM, we access a larger number of modes in the linear and quasi-linear regimes since gravitationally-induced nonlinearities are smaller. Therefore, line clustering statistics can be accurately described by perturbation theory over a wider range of scales, allowing for robust and high-precision cosmological constraints. Third, the wide redshift coverage of LIM surveys allows for efficiently breaking parameter degeneracies present in the CMB and lower-redshift probes of LSS \citep{Archidiacono:2016lnv,Obuljen:2017jiy,Lorenz:2017fgo,Sprenger:2018tdb}.

Besides the 21-cm hyperfine transition of neutral hydrogen, the rotational lines of carbon monoxide, CO, \citep{Righi:2008br,Lidz:2011dx,Breysse:2014uia,Li:2015gqa,Fonseca:2016qqw,Padmanabhan:2017ate}, and the fine structure line of ionized carbon, [CII], \citep{Gong:2011ts,Silva:2014ira,Fonseca:2016qqw,Pullen:2017ogs,Padmanabhan:2018yul} are among the most-studied target lines in the context of galaxy and star formation. More recently, some of their potential in constraining cosmology has also been explored \citep{Karkare:2018sar,Creque-Sarbinowski:2018ebl,MoradinezhadDizgah:2018zrs,MoradinezhadDizgah:2018lac,Liu:2020izx,Gong:2020lim,Bernal:2020lkd,Bernal:2021ylz}. The first detections of 21cm \citep{Masui:2012zc}, CO \citep{Keating:2015qva,Keating:2016pka,Keating:2020wlx} and [CII] \citep{Pullen:2017ogs} have amplified this growing interest in LIM. Current planned surveys, such as COMAP \citep{Li:2015gqa}, CCAT-Prime \citep{Aravena:2019tye}, CONCERTO \citep{Lagache_2018}, and TIME \citep{doi:10.1117/12.2057207} are expected to provide first robust detections of the CO/[CII] clustering power spectra. These data, however, will have limited utility in constraining cosmology. Theoretical guidance for design of wide-field LIM surveys, capable of reaching the required target sensitivities on various cosmological parameters, is therefore essential. 

In this paper we explore constraining neutrino properties using ground-based millimeter-wave LIM observations, in particular focusing on next-generation instrument configurations that could feasibly be deployed in the next decade.  New detector technologies are now being demonstrated that could provide the order-of-magnitude sensitivity improvements over current-generation experiments at reasonable cost.  We therefore explore the constraining power over a wide range of experimental sensitivities which encompass the possible experiments that could be fielded.  We forecast the expected uncertainties on $M_\nu$ and $N_{\rm eff}$ as a function of survey cost (parameterized as a product of spectrometer count and observing time) when only a 1-parameter extension of $\Lambda$CDM is considered, as well as when multiple degenerate, beyond-$\Lambda$CDM parameters are varied.

The rest of the paper is organized as follows. We review the physical effects of neutrino properties on LSS in Section~\ref{sec:rev}. We then describe the model of the line intensity power spectrum in Section~\ref{sec:linePS}, and outline the instrument and survey specifications in Section~\ref{sec:design}. After describing the details of our analysis methodology in Section~\ref{sec:Fisher}, we present our results in Section~\ref{sec:results}, and conclude in Section~\ref{sec:conclusion}. Supplementary information is provided in two appendices. In Appendix~\ref{app:binning} we give details of redshift binning and instrument noise, and in Appendix~\ref{app:EUCLID_Comp} we compare forecasted parameter constraints from LIM with those from Euclid, showing 2D marginalized errors and reporting the constraints on all model parameters. \vspace{-.05in} \\

\section{Imprints of neutrinos \\ on large-scale structure}\label{sec:rev}

Neutrinos affect cosmological observables through background and perturbation effects \citep{Lesgourgues:2018ncw}. In this section, we review various imprints of the effective number of neutrinos (light relics more generally) and their mass on LSS, highlighted in the existing literature. We also emphasize some unique advantages of LIM in shedding light on neutrino properties.

\subsection{Effective Number of Light Relics}

Keeping the redshift of matter-radiation equality and baryon density fixed, the primary background effects of increasing $N_{\rm eff}$ on LSS are an enhancement of the matter power spectrum on small scales and a damping of the baryon acoustic oscillation (BAO) amplitude, arising from the decrease of the ratio of baryons to cold dark matter (CDM). At the level of perturbations, higher $N_{\rm eff}$ shifts the BAO phase and lowers its amplitude. During the radiation-dominated era, the presence of free-streaming neutrinos significantly reduces the metric fluctuations (within the free-streaming scale), which drive oscillations in the photon-baryon fluid \citep{Hu:1995en}. Furthermore, the propagation of neutrino perturbations at the speed of light ``drags'' perturbations in the photon-baryon fluid, which propagate at the speed of sound, shifting the CMB acoustic peaks \citep{Bashinsky:2003tk}. These effects are then imprinted on baryon fluctuations prior to baryon drag time and later in the matter power spectrum through the BAO feature. The BAO phase shift, which has been measured in both the CMB and LSS \citep{Follin:2015hya, Baumann:2018qnt}, is a robust signature of light relics, difficult to mimic by changing the initial conditions or matter content, and is largely unaffected by nonlinear gravitational evolution in the late Universe \citep{Baumann:2017lmt}. However, constraints on $N_{\rm eff}$ from the BAO phase shift alone are shown to be weaker than those from the full power spectrum shape \citep{Baumann:2017gkg}. We therefore consider the full shape of the line power spectrum. 

\subsection{Sum of Neutrino Masses}

Massive neutrinos impact LSS in several ways. At the background level they change the expansion history, which alters cosmological distance scales measured by the BAO feature or Alcock-Paczynski (AP) test \citep{Zhen:2015yba}. As shown in \cite{Boyle:2017lzt}, constraints on neutrino masses from distance probes are weaker than those from the growth of structure (i.e., the perturbation effects). Furthermore, due to various parameter degeneracies (e.g., between $M_\nu$ and spatial curvature $\Omega_k$), BAO and AP constraints are significantly degraded once the parameter space is enlarged. At the level of perturbations, massive neutrinos affect the matter power spectrum in two ways.  On scales below their free-streaming scale, they do not contribute to gravitational clustering due to their large thermal velocities. Meanwhile, their background density contributes to the homogeneous expansion, slowing down the growth of clustering of matter fluctuations \citep{Bond:1980ha, Hu:1997vi}. These two effects lead to suppression of the matter power spectrum below the neutrino free-streaming scale \citep{Hu:1997mj}. On scales larger than the free-streaming scale of neutrinos when they become non-relativistic, the matter power spectrum is unaffected. Therefore, massive neutrinos render the growth rate of structure---which can be measured by redshift-space distortions (RSD)---scale-dependent. In our forecasts, we include both the background and perturbation effects by considering the full shape of the line power spectrum in redshift space, neglecting the non-linear effects of massive neutrinos  \citep{Brandbyge:2008rv,Bird:2011rb,Castorina:2015bma,Upadhye:2017hdl,Hannestad:2020rzl,Garny:2020rom,Garny:2020ilv,Chen:2020bdf,Bayer:2021kwg}.

When considering biased tracers (e.g., halos, galaxies, and line intensity), the scale-dependent growth rate of structure in the presence of massive neutrinos induces a small scale-dependence of the linear bias of the tracer \citep{LoVerde:2014pxa}\footnote{Scale-dependence of the halo bias as a result of scale-dependent growth was first studied in \cite{Hui:2007zh} and \cite{Parfrey:2010uy} in the context of modified gravity.}, which can be accurately computed using tools provided by  \cite{Munoz:2018ajr,Valcin:2019fxe}. Apart from this physical scale-dependence of the linear bias, it was shown in \cite{Villaescusa-Navarro:2013pva, Castorina:2013wga} that in the presence of massive neutrinos, defining the halo bias with respect to the total matter overdensity, $\delta_m$, results in a spurious scale-dependence of the linear bias, which can be removed if the bias is defined with respect to the CDM+baryon overdensity, $\delta_{\rm cb}$. Not accounting for this effect results in overestimating the total impact of massive neutrinos on the power spectrum of biased tracers \citep{Obuljen:2017jiy}\footnote{For instance, for the combined Euclid+HIRAX data, it was shown in \cite{Obuljen:2017jiy} that this effect can lead to degradation of the errors by roughly $30\%$.}. Therefore, we define the line bias with respect to $\delta_{\rm cb}$, and neglect the physical scale dependence of the linear bias. The latter simplification is justified since, on the scales where the effect is most prominent, the contribution of nonlinear biasing of the line intensity fluctuations (which we have neglected) most likely dominates over this effect. 

Lastly, in the presence of massive neutrinos, the change in the expansion history (resulting in a modification of the mean comoving density $\bar \rho$) and small-scale suppression of the matter power spectrum (resulting in a lower variance of matter fluctuations), also affect the halo abundance described by the halo mass function (HMF) by reducing the number of higher mass halos \citep{Castorina:2013wga, Biswas:2019uhy}. By comparing cosmological simulations with and without massive neutrinos, it has been shown that in order to obtain a universal halo mass function, replacing $\delta_m$ with $\delta_{\rm cb}$ is essential. An interesting fact relevant for LIM is that the impact of massive neutrinos on the mass function becomes more pronounced at higher redshifts. Since the line intensity fluctuations are a cumulative signal over contributions from all galaxies emitting a given line, in the halo model description, the line power spectrum is sensitive to the halo mass function. Therefore, this sensitivity offers LIM additional constraining power on neutrino masses. In particular, it is of interest to understand whether LIM across a wide $z$ range can probe the redshift-dependent modification of the HMF due to massive neutrinos, and ameliorate the degeneracy between neutrino mass and variance of fluctuations $\sigma_8$. We account for the above modification to the HMF in our forecast, but leave more detailed study of the significance of this redshift dependence to future work.

\subsection{Unique Advantages of LIM}

We now highlight the unique capabilities of LIM in constraining $N_{\rm eff}$ and $M_\nu$. Most obviously, LIM can provide significantly smaller statistical errors for both parameters than existing constraints by probing a larger comoving volume. For $N_{\rm eff}$, combining LIM and CMB data significantly improves constraints by breaking degeneracies with $\Lambda$CDM parameters and the primordial Helium fraction, $Y_{\rm He}$. The latter degeneracy is one of the main limiting factors for CMB observations, since both parameters alter the damping tail of the CMB power spectrum by changing the early-time Hubble parameter. On the contrary, as for other tracers of LSS, line intensity fluctuations are largely insensitive to $Y_{\rm He}$. For $M_\nu$ the wide redshift coverage of LIM plays a more significant role than for $N_{\rm eff}$. The reason is twofold: first, while the suppression of the small-scale matter power spectrum decreases at higher redshift, the suppression of the growth rate is more prominent at higher redshift\footnote{The latter is because at higher redshifts, the effect of neutrino free-streaming has had less time to accumulate, while the former is because once neutrinos become non-relativistic at low redshifts, they cease to affect the growth rate.}. Therefore, with a long redshift arm, LIM surveys provide a powerful means to maximally capture information from the two signatures. Second, measuring both high- and low-redshift information breaks parameter degeneracies present in the CMB and low-redshift observables. For instance, constraining the amplitude of fluctuations over a long redshift range with LIM \citep{Schmittfull:2017ffw, Yu:2018tem}, and probing LSS at the redshifts where the impact of DE is less important, lead to enhanced sensitivity to $M_\nu$. 

It should be kept in mind that when considering LIM with a single emission line, the large uncertainties on the nuisance astrophysical parameters (i.e., the mean brightness temperature and the bias of the line) can limit the constraining power since they are degenerate with cosmological parameters. As previously shown in the context of 21 cm intensity mapping, these degeneracies can be significantly alleviated in two ways. First, by modeling the line signal at higher order in perturbation theory and extending the analysis to smaller scales \citep{Castorina:2019zho}; second, by taking advantage of cross-correlations between intensity maps and other cosmological probes, such as optical galaxy surveys and CMB lensing \citep{Obuljen:2017jiy,Chen:2018qiu}. For the multi-line LIM survey we consider here, measuring cross-correlations between lines in the same survey, in addition to the auto-correlations of individual lines, provides an internal means to break the aforementioned degeneracies. In this paper we use a linear model of the line power spectrum, assume that tight priors on the line biases and mean brightness temperatures are available, and set them to their theoretically-predicted values. We leave for future work a quantitative study of the impact of these degeneracies on forecasted constraints, and the precision of the priors that the cross-correlations can provide.  \vspace{-.05in} \\

\section{The Power spectrum of \\ line intensity fluctuations \\}\label{sec:linePS}

CO is predominantly found in the dense clouds of
molecular gas (of density $\sim 10^3 \ {\rm cm}^{-3}$), 
while [CII] emission can originate from a variety of environments  \citep{Goldsmith_2012, Lagache_2018}. Both are typically tracers of the cool gas within galaxies that provides the fuel for star formation, and the strength of their emission is observed to be correlated with the star formation rates (SFRs) of galaxies \citep{Tacconi_2013,Herrera-Camus:2014qba}.

We use a simple model for the line intensity signal, commonly used in the literature, in which the mean and fluctuations of the line intensity are related to the abundance of halos that host CO- or [CII]-luminous galaxies \citep{visbal:2010rz, Gong:2011ts, Lidz:2011dx, Silva:2014ira}; see also \citealt{2016MNRAS.461...93P, 2018MNRAS.473..271V, 2018arXiv180511093P, 2018A&A...609A.130L, Yang2021} and references therein for more detailed modeling of CO and [CII] based on semi-analytical models in combination with hydrodynamical simulations.

The mean brightness temperature (typically in units of $\mu$K) at redshift $z$ is given by
\begin{equation}
\bar T_{\rm line} (z)  = \frac{c^2 p_{1,\sigma}}{2k_B \nu_{\rm obs}^2} \int_{M_{\rm min}}^{M_{\rm max}} dM \frac{dn}{dM} \frac{L_{\rm line}(M,z)}{4 \pi \mathcal{D}_{L}^{2}} \left ( \frac{dl}{d\theta} \right )^{2} \frac{dl}{d\nu},
\end{equation}
where $M_{\rm min}$ and $M_{\rm max}$ are the minimum and maximum masses of the halos that host galaxies emitting in a given line, $c$ is the speed of light, $k_B$ is the Boltzmann factor, $\nu_{\rm obs}$ is the observed frequency of the line, $dn/dM$ is the halo mass function, $L_{\rm line}$ is the specific luminosity of the line-emitting galaxy located in a halo of mass $M$ at redshift $z$, and ${\mathcal D}_L$ is the luminosity distance. The terms $dl/d\theta$ and $dl/d\nu$ reflect the conversion from units of comoving lengths, $l$, to those of the observed specific intensity: frequency, $\nu$, and angular size, $\theta$. The term $ dl/d\theta$ is equivalent to the comoving angular diameter distance and 
\begin{equation}
\frac{dl}{d\nu} = \frac{c(1+z)}{\nu_{\rm obs}H(z)}, 
\end{equation}
where $H(z)$ is the Hubble parameter at a given redshift. The parameter $p_{1,\sigma}$ accounts for scatter in the relations between SFR and specific luminosity with halo mass \citep{Li:2015gqa,Keating:2016pka} and is given by
\be\label{eq:scatter}
p_{n,\sigma} = \int_{-\infty}^\infty dx \frac{10^{n x}}{\sqrt{2\pi}\sigma_{\rm line}} e^{-x^2/2\sigma_{\rm line}^2},
\ee
with $n=1$ for the mean temperature and $n=2$ for the shot-noise contribution, as discussed below. We set the value of  $\sigma_{\rm line} = 0.37$ for both CO and [CII], corresponding to the fiducial model of \cite{Li:2015gqa} and in reasonable agreement with observational studies  \citep{Speagle2014,Carilli:2013qm,Kamenetzky_2016,Sargsyan2012}.

\begin{figure}[t]
    \flushleft\includegraphics[width=0.45 \textwidth]{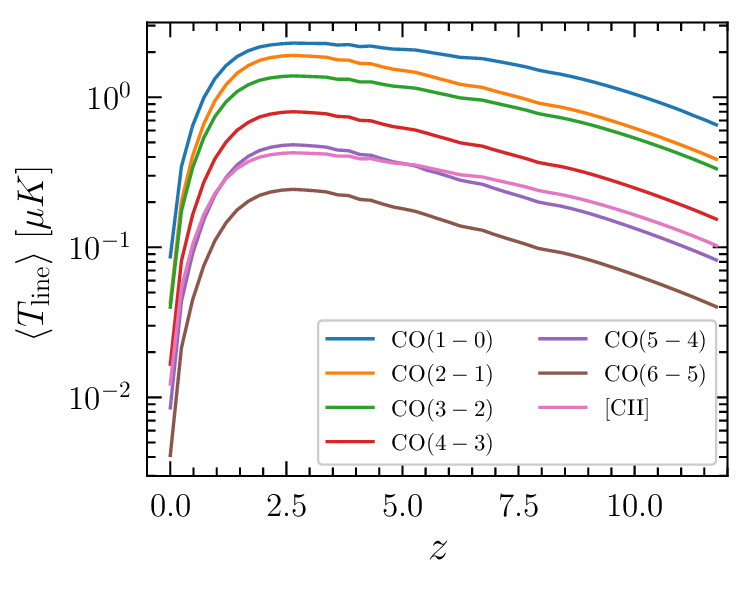}\vspace{-.1in}
    \caption{Mean brightness temperature as a function of redshift for spectral lines considered in this work.}
    \label{fig:tbar}
\end{figure}

\begin{figure*}[htbp!]
    \centering
   \flushleft \includegraphics[width=0.45 \textwidth]{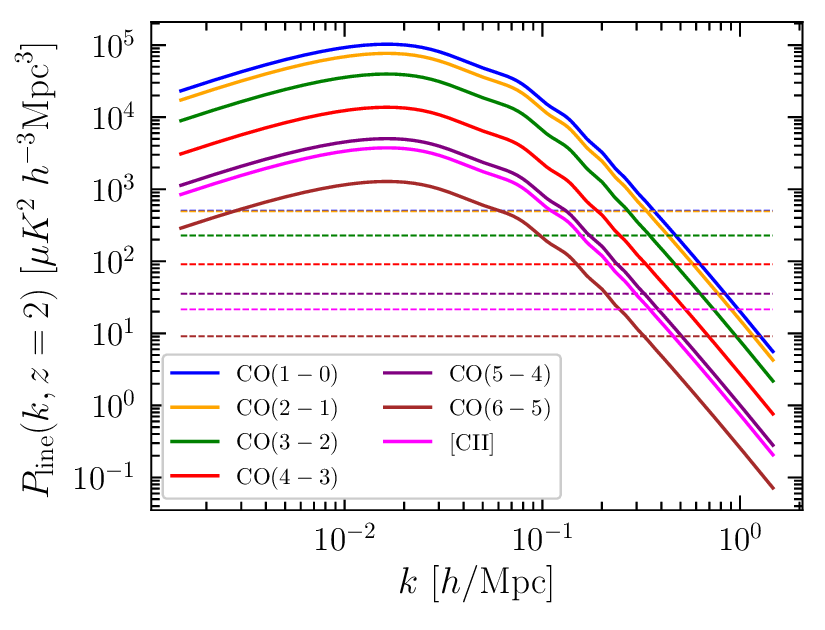}
    \hspace{.5in}\includegraphics[width=0.45 \textwidth]{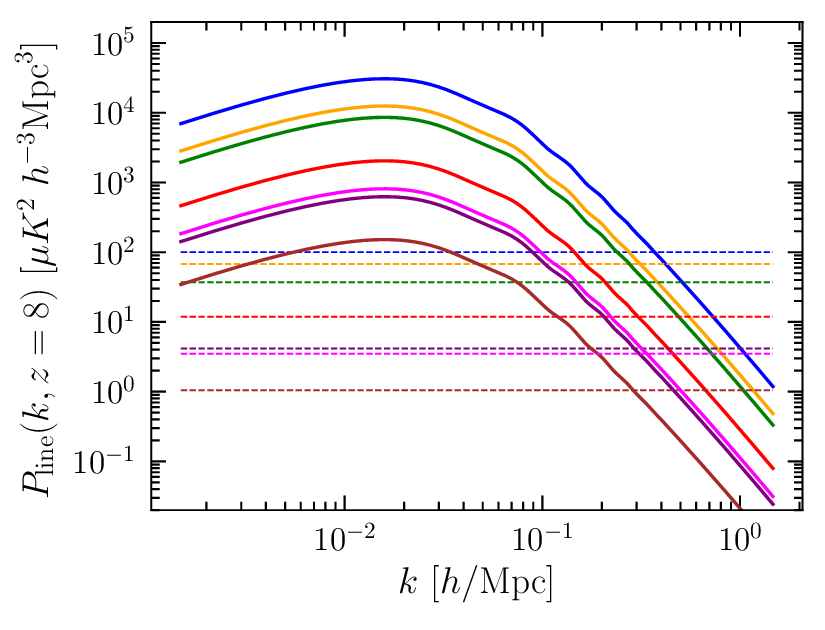}\vspace{-.1in}
    \caption{Angle-averaged clustering (solid lines) and shot noise (dotted lines) contributions to the CO/[CII] line power spectra, at $z=2$ (left) and $z=8$ (right), for $\Lambda$CDM cosmology.}
    \label{fig:line_pk}\vspace{.2in}
\end{figure*}

We model the CO luminosity by assuming scaling relations between CO and far-infrared (FIR) luminosities and between the FIR luminosity and SFR, following  \cite{Li:2015gqa}. We use the empirical fit of \cite{Behroozi:2012sp} to relate the SFR and the halo mass and redshift. At $z>8$, we extrapolate the SFR fit at lower redshift according to 
\begin{align}\label{eq:SFR_extrap}
\log  {\rm SFR}(M,z) = {\rm min}&\left[{\rm SFR}(M,z=8) + 0.2943(z-8), \right. \nonumber \\
					&\left. 3.3847-(0.2413z)\right].
\end{align}
The SFR (in units of $M_\odot {\rm yr}^{-1}$) is related to the total infrared luminosity $L_{\rm IR}$ (in units of $L_\odot$) via the Kennicut relation  \citep{Kennicutt:1998zb} of the form
\begin{equation}
{\rm SFR}(M,z) = \delta_{\rm MF} \times 10^{-10} L_{\rm IR},
\end{equation}
where the normalization depends on the assumptions of the initial mass function, the duration of star formation, etc. As in \cite{Behroozi:2012sp} and \cite{Li:2015gqa}, we take $\delta_{\rm MF} =1$. The FIR luminosity $L_{\rm IR}$ (in units of $L_\odot$) is then related to the CO line luminosity $L'_{\rm CO}$ (in units of ${\rm K} \ {\rm km} \ {\rm s}^{-1} \ {\rm pc}^2$), through a power-law fit of the form
\begin{equation}\label{eq:CO_lum}
\log L_{\rm IR} = \alpha \log  L'_{\rm CO} + \beta.
\end{equation}
We use the results of \cite{Kamenetzky_2016}, which provides empirical fits for the above relation for the CO rotational ladder using Herschel SPIRE Fourier Transform Spectrometer data. The CO line luminosity can then be expressed (in units of $L_{\odot}$) via the following expression: 
\be
L_{\rm CO(J\rightarrow J-1)} = 4.9 \times 10^{-5} J^3 \ L'_{\rm CO(J\rightarrow J-1)}.
\ee

In modeling the luminosity of [CII]-luminous galaxies, we use the results of \cite{Silva:2014ira}, which relate the [CII] luminosity to the SFR via a power-law scaling relationship, where
\be \label{eq:[CII]_lum}
\log L_{\rm [CII]} = a_{\rm L_{\rm [CII]}} \times \log {\rm SFR}(M,z) +b_{\rm L_{\rm [CII]}}.
\ee
We set the values of $a_{\rm L_{\rm [CII]}}= 0.8475$ and $b_{\rm L_{\rm [CII]}} =  7.2203$ (their model ${\bf m}_1$), which corresponds to the fit to high-redshift galaxies by \cite{DeLooze:2014dta}. We adopt the same SFR as for the CO lines. In Fig.~\ref{fig:tbar}, we show the mean brightness temperature as a function of redshift for [CII] and the first six rotational lines of CO (see also  Fig.~\ref{fig:atmosphere} for the mean brightness temperature in frequency space).

The total observed power spectrum of fluctuations in a given line, i.e., the signal of interest (s), has three contributions: clustering, shot and instrumental noise,
\be
P_{\rm tot}(k,\mu,z) = P_{\rm clust}^{\rm s}(k,\mu,z) + P_{\rm shot}^{\rm s}(z) + P_N.
\ee
Here $\mu$ is the cosine of the angle between a given wavenumber and the line-of-sight direction. The clustering contribution (typically in units of $\mu$K$^2$ Mpc$^{-3}h^3$) is anisotropic due to redshift-space distortions (RSD) and the Alcock-Paczynski (AP) effect, and on large scales can be modeled as
\begin{align}\label{eq:line_clust}
P_{\rm clust}^{\rm s}(k,\mu,z) &= \frac{H_{\rm true}(z)}{H_{\rm ref}(z)}  \left[\frac{D_{A,{\rm ref}}(z)}{D_{A,{\rm true}}(z)}\right]^2 \notag \\
&\hspace{-.2in}\times \left[1+\mu_{\rm true}^2\beta(k_{\rm true},z)\right]^2 {\rm exp}\left(-\frac{k_{\rm true}^2 \mu_{\rm true}^2 \sigma_v^2}{H^2(z)}\right),\notag \\
&\hspace{-.2in}\times \left[\bar T_{\rm line}(z)\right]^2 b_{\rm line}^2(z) P_0(k_{\rm true},z),
\end{align}
where $\beta= f/b_{\rm line}$, and $f = d\ln D(k,z)/d\ln a$ is the growth rate of structure with $D(k,z)$ the growth factor and $a$ the scale factor. Converting the
measured redshifts and angular positions to three-dimensional comoving coordinates requires making an assumption of a ``reference'' cosmology. The AP effect refers to the apparent anisotropy in the observed power spectrum induced by the discrepancy between the reference cosmology and the ``true'' cosmology, which distorts the radial and transverse distances differently. In Eq.~\eqref{eq:line_clust} this is modeled by an overall volume re-scaling factor in the first line, and by evaluating the second and third lines at the wavenumber and angles in the true cosmology ($k_{\rm true}, \mu_{\rm true}$), which are related to those in the  reference cosmology ($k, \mu$) by
\begin{align}\label{eq:trueref}
k_{\rm true} &= k \left[(1-\mu^2) \frac{D_{A,{\rm ref}}^2(z)}{D_{A,{\rm true}}^2(z)} + \mu^2 \frac{H_{\rm true}^2(z)}{H_{\rm ref}^2(z)}\right]^{1/2},  \nonumber \\
\mu_{\rm true} &= \frac{k \mu }{k_{\rm true}} \ \frac{H_{\rm true}(z)}{H_{\rm ref}(z)}.
\end{align}

The factors in the second line of Eq.~\eqref{eq:line_clust} account for RSD, i.e., distortions induced by the peculiar velocities of galaxies emitting a given line. The first is the linear Kaiser term accounting for enhancement of power on large scales, while the second is the suppression of power on small scales, i.e., the Finger of God (FoG) effect. The effect of intrinsic line width of individual emitters, over which the emission is smeared out (as discussed in \citealt{Keating:2020wlx}), can be described similarly to the FoG suppression. Therefore, we have
\begin{equation}\label{eq:disp}
    \sigma_v^2 = (1+z)^2 \left[\frac{\sigma^2_{\rm FoG}(z)}{2} + c^2 \sigma_z^2\right],
\end{equation}
where 
\begin{equation}
    \sigma_{\rm FoG}(z) = \sigma_{\rm FoG,0} \sqrt{1+z}.
\end{equation}
For both CO and [CII], we assume that the bulk of the emitters have line widths of $300 \ \textrm{km}\ \textrm{sec}^{-1}$, approximately matching those for typically-observed CO-bright galaxies at high redshift \citep{Tacconi:2013gf}, and corresponding to a value of $\sigma_z = 0.001(1+z)$. We vary $\sigma_{\rm FoG,0}$ as a free parameter in our forecasts. 

Finally, the third line of Eq.~\eqref{eq:line_clust} is the real-space clustering power spectrum with $P_0(k,z)$ the linear dark matter power spectrum, and $b_{\rm line}(z)$ the luminosity-weighted bias of the line intensity, related to bias of halos with mass $M$ at redshift $z$, $b_h(M,z)$ as 
\begin{equation}\label{eq:line_bias}
b_{\rm line}(z) = \frac{\int_{M_{\rm min}}^{M_{\rm max}} dM  \ \frac{dn}{dM} \  b_h(M,z) L_{\rm line}(M,z)  
}{\int_{M_{\rm min}}^{M_{\rm max}} dM \ \frac{dn}{dM} \ L_{\rm line}(M,z)}.
\end{equation}

In the Poisson limit, the shot-noise contribution to the line power spectrum arising from discrete nature of sources of line emission is given by 
\begin{align}\label{eq:ps_shot}
P_{\rm shot}^{\rm s}(z) = \frac{c^4 p_{2,\sigma}}{4 k_B^2 \nu_{\rm obs}^4} 
& \int_{M_{\rm min}}^{M_{\rm max}} dM  \ \frac{dn}{dM} \notag \\ &\times{\left[\frac{L_{\rm line}(M,z)}{4 \pi \mathcal D_L^2} 
 \left ( \frac{dl}{d\theta} \right )^{2} \frac{dl}{d\nu} \right ]}^2,
\end{align} 
where $p_{2,\sigma}$ is given by Eq.~\eqref{eq:scatter} for $n=2$.

\begin{figure}[t]
    \flushleft\includegraphics[width=0.45\textwidth]{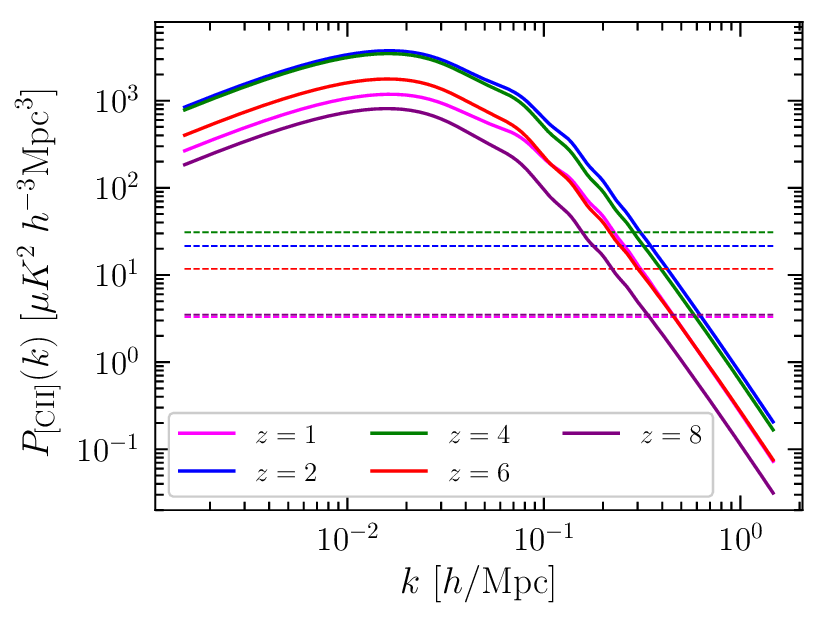}\vspace{-.1in}
    \caption{Redshift evolution of the angle-averaged [CII] power spectrum for $\Lambda$CDM cosmology.} 
    \label{fig:pk_CII_z}\vspace{.1in}
\end{figure}
To illustrate the relative amplitudes of the lines considered in our forecasts, in Fig.~\ref{fig:line_pk} we plot the clustering (solid lines) and shot components (dotted lines) of the CO/[CII] power spectra at $z=2$ (left) and $z=8$ (right). For the clustering contribution we show the monopole, averaged over the angle $\mu$. Note the change in relative amplitude of [CII] and CO(5-4) between the two redshifts. This is in agreement with the cross-over seen at $z\sim 5$ in Fig.~\ref{fig:tbar}, with [CII] becoming brighter than CO(5-4) at higher redshift. To better illustrate the redshift evolution of [CII], we show the angle-averaged [CII] power spectrum in Fig.~\ref{fig:pk_CII_z}. The amplitude first increases and then decreases with increasing redshift. This behavior is largely driven by star-formation history, although the redshift evolution of the line bias, matter fluctuations, and growth rate also have an impact. 

Before closing this section, we note two aspects of the line power spectrum model used in this paper that will be improved in future, more realistic forecasts. First, for the clustering component in Eq.~\eqref{eq:line_clust}, it is assumed that the matter fluctuations, the biasing relation of the line intensity, and the RSD can be described linearly. The linear model is clearly only valid on very large scales, and will be extended to include one-loop contributions. Second, the shot noise is assumed to be described by the Poisson approximation. Since on large scales the halo exclusion and small-scale nonlinearities are expected to produce sub/super-Poissonian shot noise, the model will treat this correction as a nuisance parameter to be marginalized over. In the context of CO/[CII] LIM, these two aspects were recently studied in Ref.~\citep{MoradinezhadDizgah:2021dei}, where the model predictions were tested against simulated intensity maps. Including these ingredients in the Fisher forecasts is likely to weaken the reported constraints on cosmological parameters, as has been shown in the context of intensity mapping of neutral hydrogen \citep{Sailer:2021yzm}. 

\section{Instrument Specifications, \\ 
        Survey Design, and Noise}\label{sec:design}

In this section we discuss the specifications of a hypothetical next-generation mm-wave LIM survey, and how those specifications translate into sensitivity estimates used in our forecasts.

In a LIM experiment, an image cube is generated by measuring the specific intensity at several sky positions, $(l,m)$, and frequencies, $\nu$. A Fourier transform produces a 3D power spectrum, which can be further averaged down to a 1D spectrum like those seen in Figs.~\ref{fig:line_pk} and \ref{fig:pk_CII_z} under the assumption of the cosmological principle. The noise on the individual modes of the power spectrum, $P_{\rm N}$, can be related to the per-voxel noise, $\sigma_{\rm N}$, of the original image cube:
\begin{equation}\label{eqn:noisepermode}
    P_{\rm N} = \sigma_{N}^{2}V_{\rm vox}.
\end{equation}
$V_{\rm vox}$ is the volume of individual voxels within the image cube, which can be further expressed as
\begin{equation}\label{eqn:voxelvolume}
    V_{\rm vox} = \Omega_{\rm vox}\delta\nu \left( \frac{dl}{d\theta} \right )^{2} \frac{dl}{d\nu}.
\end{equation}
Here $\Omega_{\rm vox}$ and $\delta\nu$ are the solid angle and bandwidth covered by a single voxel, respectively.

The sensitivity of our hypothetical survey is parameterized by \textit{spectrometer-hours}, $\tau_{\rm sh}$, as a proxy for the ``effort level'' of an experiment. If the survey area is $\Omega_{\rm s}$, such that the number of independent pointings is given by $\Omega_{\rm s}/\Omega_{\rm vox}$, then we can express Eq.~\eqref{eqn:noisepermode} as
\begin{equation}
    P_{\rm N} = \frac{\Omega_{\rm s} \sigma_{\rm NET}^{2} \delta\nu }{{\eta_{opt}^{2}\tau_{\rm sh}}} \left( \frac{dl}{d\theta} \right )^{2} \frac{dl}{d\nu},
\end{equation}
where $\sigma_{\rm NET}$ is the noise-equivalent temperature (NET) of the detector (units of K$\cdot\sqrt{\rm s}$), and $\eta_{\rm opt}$ is the optical efficiency of the instrument.

\begin{figure*}[t]
\includegraphics[width=0.95\textwidth]{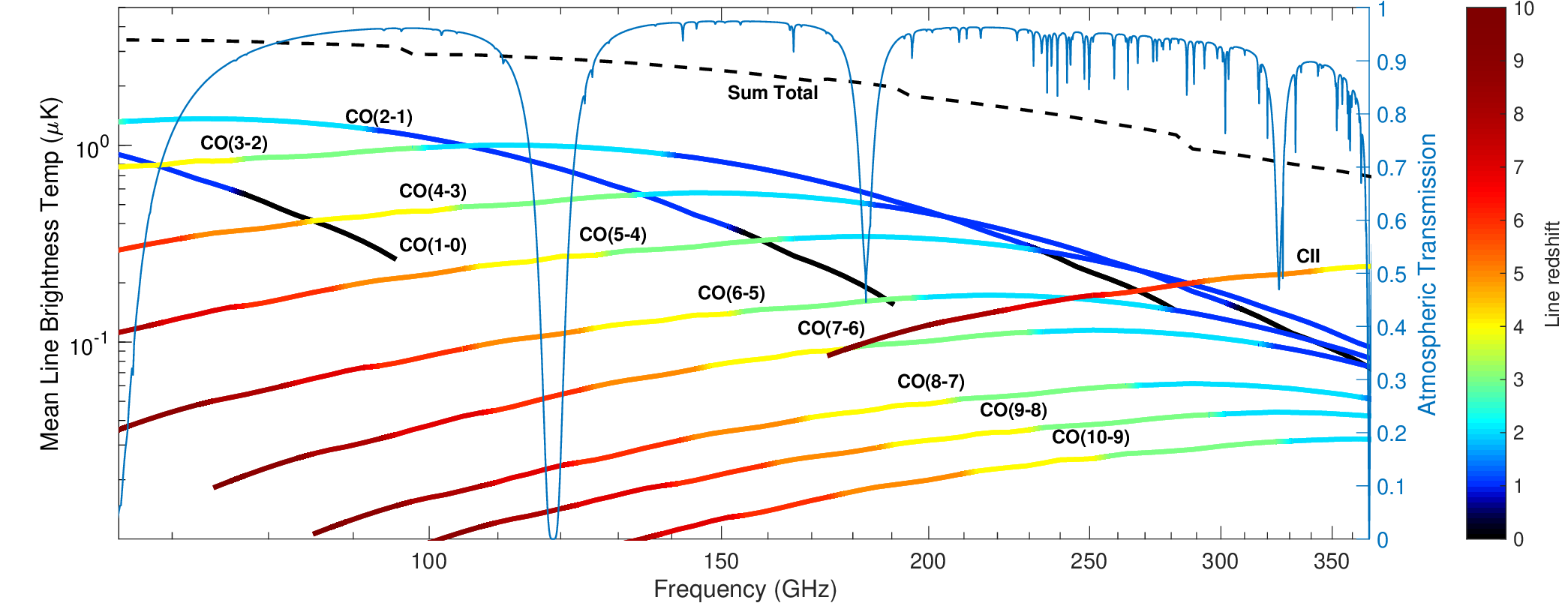}
\caption{ Spectral lines detectable by a ground-based survey. Shown above are model predictions for the brightness temperature of individual lines, their sum total (in dashed black), and atmospheric transmission (in light thin blue). The colors correspond to observed redshifts of the lines. The shown signal strengths adopt line luminosities scaled from the IR luminosity based on observational relations. Current constraints on these line ratios are uncertain by up to one order of magnitude.}
\label{fig:atmosphere}\vspace{.2in}
\end{figure*}
For our analysis, we consider ground-based observations from an accessible observing site with excellent proven mm-wave observing conditions, such as the South Pole or the Atacama desert. Fig.~\ref{fig:atmosphere} shows the atmospheric transmission as a function of frequency for median South Pole winter conditions, calculated using the \textit{am} software \citep{paine2019}. In our projections, each individual spectrometer is sensitive to the full frequency range\footnote{By `full frequency range', we mean that each spectrometer is capable of measuring all individual spectral channels simultaneous. This is in contrast to e.g., Fabry-Perot interferometers or Fourier transform spectrometers, which require multiple measurements at different delays to synthesize individual frequency channels, and thus have lower instantaneous throughput than the spectrometer considered here.} between 80--310 GHz, encompassing the typical ground-based CMB atmospheric ``windows'' at 1, 2, and 3 mm; similar wide-bandwidth observations in a single receiver have been made with e.g., SPT-3G \citep{Anderson:2019dhb}. We assume each spectrometer has a spectral resolution of $R= \nu/\Delta \nu = 300$, equivalent to what is now being demonstrated in wideband mm-wave spectrometers \citep{redford2018, karkare2020}. The instrument is assumed to have a 10 m aperture, similar to the South Pole Telescope.  We assume an overall optical efficiency of 25\%, typical of current CMB experiments, and that each spectrometer detects both polarizations.  

In the last two decades, mm-wave detectors have been demonstrated with device noise contributions that are well below atmospheric noise at these sites, even for narrow spectroscopic channels.  We therefore make the assumption that each spectrometer's noise is dominated by atmospheric emission, with a secondary contribution from the emissivity of the telescope reflector surface. We define $Q_{\rm tot}$ to be the power received by a single channel of the detector:
\begin{equation}
    Q_{\rm tot} = Q_{\rm sky} + Q_{\rm tel},
\end{equation}
where $Q_{\rm sky}$ and $Q_{\rm tel}$ are the noise power seen from the sky and the telescope respectively. These individual contributions, under the assumption of single-mode, diffraction-limited optics, can be further expressed as
\begin{equation}\label{eqn:detpower}
    Q = \eta_{\rm opt} \lambda_{\rm obs}^{2} \epsilon(\nu) \Delta\nu  B(\nu, T)   .
\end{equation}
Here, $\lambda_{\rm obs}$ is the observed wavelength, $\epsilon$ is the emissivity, and $B$ is the Planck function, which varies as a function of frequency, $\nu$, and temperature, $T$. We take the temperature to be 250 K for both the telescope and the atmosphere. The emissivity of the telescope is fixed at $\epsilon_{\rm tel}=0.01$, while the emissivity of the atmosphere $\epsilon_{\rm atm}$ is calculated as a function of frequency, varying between $0.05-1$ over the 80--310 GHz range. We can  express the NET as
\begin{equation}\label{eqn:sigma_net}
    \sigma_{\rm NET} =
    \frac{\sqrt{2Q_{\rm tot} h \nu + (Q^{2}_{\rm tot} / \Delta\nu)}}
    {2k_{\rm B}\eta_{\rm opt}\Delta\nu (1-\epsilon_{tel})(1-\epsilon_{\rm atm})},
\end{equation}
where $h$ is the Planck constant. Typical NET values for our hypothetical instrument are of order ${\sim} 1$ mK$\cdot\sqrt{\rm s}$ over the frequency range of interest. The NET can change substantially over the redshift window of a given line, particularly around the 118 GHz oxygen line or the 183 GHz water line. An optimal accounting for such variations in instrument noise are likely to be survey- and instrument-dependent, and is therefore beyond the scope of the work here. As a first-order approximation, we therefore take the median NET over a redshift window when evaluating Eq.~\eqref{eqn:noisepermode} for our Fisher analysis, and exclude frequency channels with NETs more than a factor of $\sqrt{2}$ greater than this median. The fraction of excluded channels is factored into the effective volume for each line and redshift combination (the use of which is discussed in Section~\ref{sec:Fisher}). 

As the optimal survey size varies as a function of survey sensitivity,  for a given value of spectrometer-hours we find the sky fraction that provides the tightest parameter constraints.  This is evaluated over $17 \ \mathrm{to} \ 17000 \  \mathrm{deg}^2$.  The minimum roughly corresponds to the smallest area that can be efficiently surveyed with a wide field-of-view mm-wave camera, while the maximum has recently been demonstrated by ACT from the Atacama desert \citep{Naess:2020wgi}.

In our forecasts we vary spectrometer-hours over a wide range, starting with first-detection experiments and extending to larger-scale surveys that could be fielded in the next ten years. Current-generation instruments feature $\sim 50$ spectrometers \citep{doi:10.1117/12.2057207} and are capable of completing surveys of order $10^{5}$ spectrometer-hours. We therefore use this as a lower bound of the range considered here. 

Current wide-bandwidth spectrometers either use a free-space diffraction grating that limits the number of detectors that can be placed in a cold volume, or use a Fourier Transform spectrometer or Fabry-Perot interferometer, which do not simultaneously detect all frequencies independently. However, on-chip spectrometer technology---in which all frequencies are simultaneously and individually measured---can substantially improve sensitivity. Current examples include SuperSpec/DESHIMA (filter-bank; \citealt{Shirokoff_2012, endo2019}) and $\mu$-Spec (grating; \citealt{cataldo2012}), all of which perform background-limited, wide-bandwidth spectroscopy on a few cm$^2$ of silicon.  This technology will soon lead to filled spectroscopic focal planes similar in format to those used by CMB experiments.  In several years, $\sim 400$ spectrometers could potentially be fielded in a single optics tube for planned multi-tube receivers.  Future instruments could host anywhere from 7 (CCAT-prime; \citealt{Aravena:2019tye}) to 85 optics tubes (CMB-S4 LAT; \citealt{Abazajian:2019eic}).  CMB facilities at the South Pole and in Chile routinely accumulate several thousand hours of integration time per year, and survey operations can extend for five years or more.  We therefore forecast constraints for a wide range of spectrometer-hours, extending up to $4\times 10^9$.

As in \cite{MoradinezhadDizgah:2018lac}, we define an effective instrumental noise,
\be
\tilde P_{\rm N}(k,\mu,z) =  \alpha^{-1} _{\rm max}(k,\mu)\alpha^{-1} _{\rm min}(k,\mu)P_{\rm N},
\ee 
to account for attenuation of the signal due to the finite resolution of the instrument ($\alpha_{\rm max}$) and the finite redshift and angular coverage of the survey ($\alpha_{\rm min}$). The two attenuation factors are defined in terms of the largest and smallest recoverable modes in parallel and perpendicular to line-of-sight directions, 
\begin{align}\label{eqn:atten_kmin}
\alpha_{\rm min}(k,\mu) &= \alpha_{\rm min}(k_{\perp},k_{\parallel}) \\ 
&= \left (1-e^{-k_{\perp}^{2}/(k_{\perp,{\rm min}}/2)^{2}}\right) \left (1-e^{-k_{\parallel}^{2}/(k_{\parallel,{\rm min}}/2)^{2}} \right), \notag
\end{align}
\vspace{-.15in}
\begin{equation}\label{eqn:atten_kmax}
\alpha_{\rm max}(k,\mu) = \alpha_{\rm max}(k_{\perp},k_{\parallel}) = e^{-(k_{\perp}{^2}/k_{\perp,{\rm max}}^{2}+k_{\parallel}{^2}/k_{\parallel,{\rm max}}^{2})},
\end{equation}
where 
\be
k_\parallel = k\mu  \qquad k_\perp = k \sqrt{1-\mu^2}.
\ee
The smallest recoverable modes are given by
\begin{equation}\label{eqn:kmin_par}
k_{{\rm max},\parallel} \approx  2\pi \left [ \delta\nu  \frac{dl}{d\nu}\right ] ^{-1},
\end{equation}
\begin{equation}\label{eqn:kmin_perp}
k_{{\rm max},\perp} \approx 2\pi \left [ \frac{c/\nu_{obs}}{D_{\rm ant}} \frac{dl}{d\theta}\right ]^{-1},
\end{equation}
where $D_{\rm ant}$ is the diameter of the aperture and $\delta \nu$ is the spectral resolution of the instrument. The largest recoverable modes are given by
\begin{equation}\label{eq:kpar_min}
k_{\parallel,{\rm min}} = 2\pi \eta_{\rm min} \left [ \nu_{obs} \frac{dl}{d\nu} \right ] ^{-1},
\end{equation}
\begin{equation}\label{eq:kperp_min}
k_{\perp,{\rm min}}\approx2\pi \left[2\sin\left (\theta_{\rm max}/2\right )\frac{dl}{d\theta} \right ]^{-1},
\end{equation}
where $\theta_{\rm max} = \sqrt{\Omega_{\rm survey}}$ is determined by the angular coverage of the survey, and $\eta_{\rm min}$ is set by the redshift coverage of the survey. We set $D_{\rm ant} = 10 \ \rm m$, $\nu_{\rm obs}/\delta\nu = 300$, and $\eta_{\rm min} = 3$, the latter of which is set by the frequency distance between high-opacity telluric lines (typically arising from oxygen and water, occurring at approximately 60, 118, 183, and 325 GHz in the millimeter-wave atmospheric window). Although in principle it may be possible to probe modes beyond this limit, this would require more sophisticated accounting for the effect of atmospheric windowing, and we conservatively assume that such modes are not practically accessible.

We note that our choice of $D_{\rm ant}$ and $\delta\nu$ are based on existing instrument parameters, although since we impose a conservative cutoff on the smallest-scale modes included in our forecasts ($k\gtrsim0.3\ h\,\textrm{Mpc}^{-1}$; see Section~\ref{sec:Fisher}), our results are expected to be relatively insensitive to these choices.  \vspace{-.05in} \\

\section{Analysis Methodology}\label{sec:Fisher}

We use the Fisher matrix formalism to perform parameter forecasts. For a given emission line, we split the survey into redshift bins with mean redshifts $z_i$ and widths of 0.1 dex to account for the cosmic evolution of the line-emitting population, as well as variations in the instrument noise caused by the frequency-dependent atmospheric transmission (as discussed in Section~\ref{sec:design}). Neglecting correlations between redshift bins, for each emission line $\rm x$, the total Fisher matrix is the sum of Fisher matrices of individual redshift bins,
\be
F_{\alpha \beta}^{\rm x} = \sum_i F_{\alpha\beta}^{{\rm x},i}
\ee
with the Fisher matrix in the $i^{\text{th}}$ redshift bin given by
\begin{align}\label{eq:Fisher_single}
F_{\alpha\beta}^{{\rm x},i} &=  V_i \int_{-1}^1\int_{k_{\rm min}}^ {k_{\rm max}}  \frac{ k^2 {\rm d}k \ {\rm d}\mu  }{8 \pi^2} {\rm Var}^{-1}[P_{\rm obs}^{\rm x}(k,\mu,z)] \nonumber \\
&\times \frac{\partial P_{\rm clust}^{\rm x}(k,\mu,z_i)}{\partial { \lambda}_\alpha } \frac{\partial P_{\rm clust}^{\rm x}(k,\mu,z_i)}{\partial {\lambda}_\beta } ,
\end{align}
where $\lambda$ are the model parameters that are varied. $V_i$ is the volume of $i^{\text{th}}$ redshift bin extended between $z_{\rm min}$ and $z_{\rm max}$ and is proportional to sky coverage of the survey, $f_{\rm sky}$, while ${\rm Var}^{-1}[P_{\rm line}^{\rm x}]$ is the variance of the line power spectrum. The median redshifts of each bin and the corresponding instrument noise are given in Table~\ref{tab:noise_per_redshift_bin} of Appendix~\ref{app:binning}. The total Fisher matrix for all lines considered (i.e., CO from  J=2-1  to  J=6-5 and [CII]) is given by the sum of the Fisher matrices of individual lines, neglecting their cross correlations, $F_{\alpha \beta}^{\rm tot} = \sum_{\rm x} F^{\rm x}_{\alpha\beta}$.

We set the fiducial $\Lambda$CDM  parameter values to those from Planck 2018 data\footnote{Specifically, we use \textsf{ base\_plikHM\_TTTEEE\_lowl\_lowE}.} \citep{Aghanim:2018eyx}: ${\rm ln} (10^{10}A_s) = 3.0447, n_s = 0.96589, h = 0.6732,\Omega_{\rm cdm} =0.2654,  \Omega_b = 0.04945$. For extension parameters we set $N_{\rm eff} = 3.046, M_\nu = 0.06, Y_{\rm He} = 0.245398, w_0 = -1, w_a = 0$. For all forecasts we assume three degenerate massive neutrino species\footnote{As first pointed out in \cite{Lesgourgues:2004ps}, considering 3 non-degenerate neutrinos and setting the mass of one (assuming an inverted hierarchy) or two (assuming a normal hierarchy) of them to zero, leads to observables significantly different from the true ones.}. This choice is motivated by recent results showing that the assumption of degenerate neutrinos is sufficiently accurate for current cosmological observations \citep{RoyChoudhury:2019hls}. It is also worth noting that while high-precision measurement of $M_\nu$ by future surveys can rule out the inverted hierarchy, even future cosmological data alone will most likely not be able to directly detect the neutrino mass hierarchy \citep{Archidiacono:2020dvx}\footnote{The weak evidence for the normal hierarchy in current data is almost entirely driven by the prior volume; the inverted hierarchy significantly reduces the size of parameter space \citep{Hall:2012kg}.}. 

We choose the fiducial value of $\sigma_{{\rm FOG},0} = \ 250 \ {\rm km}/{\rm s}$, and fix the line bias and mean brightness temperature to their theoretical values, accounting for their cosmology-dependence. We show results from LIM alone and in combination with Planck. For the latter, for each cosmology we use the corresponding publicly available Planck 2018 MCMC chains\footnote{\url{http://pla.esac.esa.int/pla/\#cosmology}} to compute the parameter covariances and the Fisher matrix, marginalizing over optical depth. We assume the CMB and LIM Fisher matrices to be independent and add them to obtain the joint constraints. We use the CLASS code \citep{Blas:2011rf}\footnote{\url{http://class-code.net}} to compute the matter power spectrum.

We compute $k_{\rm min}$ from the largest recoverable modes in the parallel and perpendicular to line-of-sight directions given in Eqs.~(\ref{eqn:kmin_par}, \ref{eqn:kmin_perp}):
\be
k_{\rm min} = \sqrt{k_{\parallel, {\rm min}}^2 + k_{\perp, {\rm min}}^2}.
\ee
We set $k_{\rm max} = 0.15$ ${\rm Mpc}^{-1}h$ at $z=0$. At higher redshifts, we obtain $k_{\rm max}$ by finding the value that satisfies the following condition for the variance of the linear matter density field:
\be
\sigma^2(z) = \int_{k_{\rm min}(z)}^{k_{\rm max}(z)} \frac{d^3 k}{(2\pi)^3} P_0(k,z) = \sigma^2(z=0).
\ee 
This choice ensures that at a given redshift, we are in the regime where fluctuations in the matter density are in the nearly-linear regime where perturbation theory is valid. We further impose a conservative upper bound of $k_{\rm max} \leq 0.3 \ {\rm Mpc}^{-1}h$ to ensure that our assumptions of linear bias and linear RSD (linear Kaiser term) are valid.   
Fisher matrices are evaluated for a given set of choices for $f_{\rm sky}$ and spectrometer-hours, combining constraints across all line species and redshift windows.

\subsection{Impact of Interloper Lines}

In a LIM survey, the voxel volume is defined in terms of the angular and frequency resolutions of the instrument, Eq.~\eqref{eqn:voxelvolume}. At a given redshift, in addition to the line of interest (i.e., the signal), lines emitted from other redshifts projected onto the same observed frequency can be confused with the target signal in a given voxel. These contributions are referred to as interloper lines and can be accounted for as an additional source of noise in the measurement of the LIM power spectrum \citep{Cheng:2016, Gong:2020lim}. Accounting for interloper lines is particularly important for measuring the power spectrum at higher redshifts, as the interlopers at longer wavelengths (lower redshifts) can significantly contaminate the signal of interest. The difference in the redshifts of the signal and interlopers affects perpendicular and parallel to line-of-sight distances differently. Therefore, the power spectra of interlopers observed at the signal's frequency are anisotropic even in the absence of RSD and the AP effect. Furthermore, in computing the power spectra of the projected interlopers, we must convert the comoving volume of the interlopers at their emission redshifts to the redshift of the signal at which they are observed. 

Putting all this together, accounting for $N$ interloper lines, the variance of the line signal power spectrum is given by 
\begin{align}\label{eq:var}
{\rm Var}&[P_{\rm obs}(k,\mu,z)] \notag \\
&\hspace{-.13in} =  \left\{P_{\rm clust}^{\rm s}(k,\mu,z) + P_{\rm shot}^{\rm s}(z) + \tilde P_N(k,\mu,z)\right. \notag \\
& \hspace{-.13in} \left. + \sum_i^N \left[A_\perp^i\right]^2 A_\parallel^i \left[P_{\rm clust}^i\big(k_\parallel^i, k_\perp^i, z_i \big) + P_{\rm shot}^i(z_i)\right]\right\}^2. 
\end{align}
The first three terms on the right hand side of Eq.~\eqref{eq:var} are the clustering and shot components of the line signal and instrumental noise, while the terms in the last line are the contributions from interloper lines. The factor of $\left[A_\perp^i\right]^2 A_\parallel^i$, accounts for the difference in comoving volume due to interloper lines emitted from $z_i$ with the volume of the line of interest at redshift $z$, given by
\be
A_\perp^i = \frac{D_A(z)}{D_A(z_i)},  \qquad \qquad A_\parallel^i = \frac{H(z_i)(1+z)}{H(z)(1+z_i)}.
\ee
The parallel and perpendicular components of the interloper wavevectors at their corresponding redshifts $z_i$ are related to those of the source at redshift $z$ by
\begin{align}
k_\parallel^i(k,\mu) &= k \ \mu \ A_\parallel^i \notag \\
k_\perp^i(k,\mu) &= k \ \sqrt{1-\mu^2} \  A_\perp^i .
\end{align}
When neglecting the impact of interloper lines, the last line in Eq.~\eqref{eq:var} is set to zero and the variance is given by the sum of the clustering and shot contributions of the line of interest, and the instrumental noise. 

Several techniques for reducing the impact of line interlopers have been proposed in the literature. This includes voxel masking \citep{Silva:2014ira,Breysse:2015baa,Sun_2018}, cross-correlation \citep{Lidz:2016lub}, spectral line deconfusion \citep{Lidz:2016lub,Cheng:2016,Gong:2020lim}, and machine learning \citep{Moriwaki_2020} or map-based deconfusion \citep{Kogut2015,Cheng:2020asz}. With realistic component separation techniques, the contribution of interlopers to the variance of the power spectrum may be less than that implied by Eq.~\eqref{eq:var}, particularly with access to multiple lines in the same redshift window, as we do in the hypothetical instrument setup specified in Section~\ref{sec:design}. Due to a lack of observational data, the efficacy of such methods in the presence of instrumentally- and atmospherically-induced spectral windowing functions is not yet known. Therefore, in our analysis we consider two scenarios, which we expect will bracket the range of expectations with future analysis techniques. In the first, we consider Eq.~\eqref{eq:var} as written (i.e., with interlopers present), and in the second, we neglect the $P^{i}$ terms, effectively evaluating the power spectra as though no interloper line species existed.

\subsection{Parameter Degeneracies}

Constraints on $N_{\rm eff}$ and $M_\nu$ can considerably degrade when enlarging the model parameter space (\citealt{Archidiacono:2016lnv,Boyle:2017lzt,Boyle:2020rxq}). Apart from degeneracies between $N_{\rm eff}$,  $M_\nu$, and $\Lambda$CDM parameters, two other well-known degeneracies are between  $N_{\rm eff}$ and $Y_{\rm He}$ in CMB data, and between $M_\nu$ and DE properties for both CMB and LSS probes at $z < 2$. To illustrate the strength of LIM in alleviating parameter degeneracies, we include six extensions to $\Lambda$CDM in our forecasts, described in Table~\ref{tab:models}. 

We consider two models of dark energy beyond the cosmological constant; one with a constant EoS, and the other a dynamical DE model assuming the Chevallier-Linder-Polarski (CLP) parameterization \citep{Chevallier:2000qy,Linder:2002et}, in which the redshift evolution of the dark energy EoS is given by
\be 
w(z) = w_0 + w_a \frac{z}{1+z};
\ee 
$w_0$ denotes the present-day value of the DE EoS. Thus, we vary $w_0$ in the first case, and $w_0$ and $w_a$ in the second case. Varying the DE EoS parameter(s) along with $M_\nu$ is shown to degrade the bounds on $M_\nu$ from a ``vanilla'' $\Lambda\rm CDM+M_\nu$ model \citep{Hannestad:2005gj,Upadhye:2017hdl,Lorenz:2017fgo,Mishra-Sharma:2018ykh,Brinckmann:2018owf, RoyChoudhury:2019hls}. However, an exception to this rule occurs when the phantom part of the DE EoS parameter space ($w(z) \leq -1$) is excluded from the analysis: due to the parameter degeneracies, the non-phantom part of the parameter space prefers lower neutrino masses than $\Lambda\rm CDM+M_\nu$ (see e.g., \citealt{Vagnozzi:2018jhn, RoyChoudhury:2018vnm}). We do not consider a non-phantom-only model in this work. 

The significant degeneracy between $M_\nu$ and the DE properties in CMB data originates from a geometric degeneracy: given that the CMB tightly constrains the comoving distance to the last-scattering surface, and that the present-day CMB photon density and the total CDM+baryon density are tightly constrained by the COBE/FIRAS temperature and CMB acoustic peaks, respectively, any change in comoving distance to the last-scattering surface due to the change in the DE EoS parameter(s) must be compensated by changes to the Hubble parameter and $M_\nu$. For LSS observables at low redshifts ($z<2$) when DE provides an important contribution to the energy density of the universe, $M_\nu$ and the DE EoS parameter(s) are correlated since they both modify the expansion rate and the growth of structure. Measurement of LSS observables at multiple redshifts is critical to alleviating this degeneracy (see e.g., \citealt{Mishra-Sharma:2018ykh}). 

\begin{table}[t]
\caption{Models considered in the Fisher forecasts.}\label{tab:models} \vspace{.1in}
\renewcommand{\arraystretch}{1.2}
\hspace{.3in}\begin{tabular}{c  c }
    \hline
    Model &  Parameters \vspace{.04in}\\
    \hline
    1 & $\Lambda{\rm CDM}+N_{\rm eff}$ \\
    2 & $\Lambda{\rm CDM}+M_\nu$  \\
    3 & $\Lambda{\rm CDM}+N_{\rm eff}+M_\nu$ \\
    4 &  $\Lambda{\rm CDM}+N_{\rm eff}+Y_{\rm He}$ \\
    5 & ${\rm CDM}+M_\nu + w_0$  \\
    6 &  ${\rm CDM}+M_\nu + w_0 + w_a$  \vspace{.04in} \\
    \hline
\end{tabular}
\end{table}

\subsection{Optimization Strategy}

For each combination of interloper scenario and choice of cosmology, we calculate constraints on the parameter of interest (either $N_{\rm eff}$ or $M_{\nu}$) while varying $f_{\rm sky}$ and spectrometer-hours, both with and without the addition of the Planck priors. We vary $f_{\rm sky}$ between 0.004 and 0.4096, stepping by factors of two, and spectrometer-hours between $2\times10^5$ and $4\times10^{9}$, stepping by factors of three, and calculate constraints for a total of 110 observational setups for each of the 24 scenarios considered in our analysis. These constraints are then interpolated using a bicubic interpolation scheme, to estimate the optimal choice of $f_{\rm sky}$ (and expected value of $\sigma(N_{\rm eff})$ or $\sigma(M_\nu)$), given a number of spectrometer-hours. \vspace{-.08in} \\

\section{Results}\label{sec:results}

In this section we present the results of our forecasts, discussing the constraints on $N_{\rm eff}$ and $M_\nu$ separately. 

\begin{figure*}[!htbp]
    \centering
    \includegraphics[width=.85\textwidth]{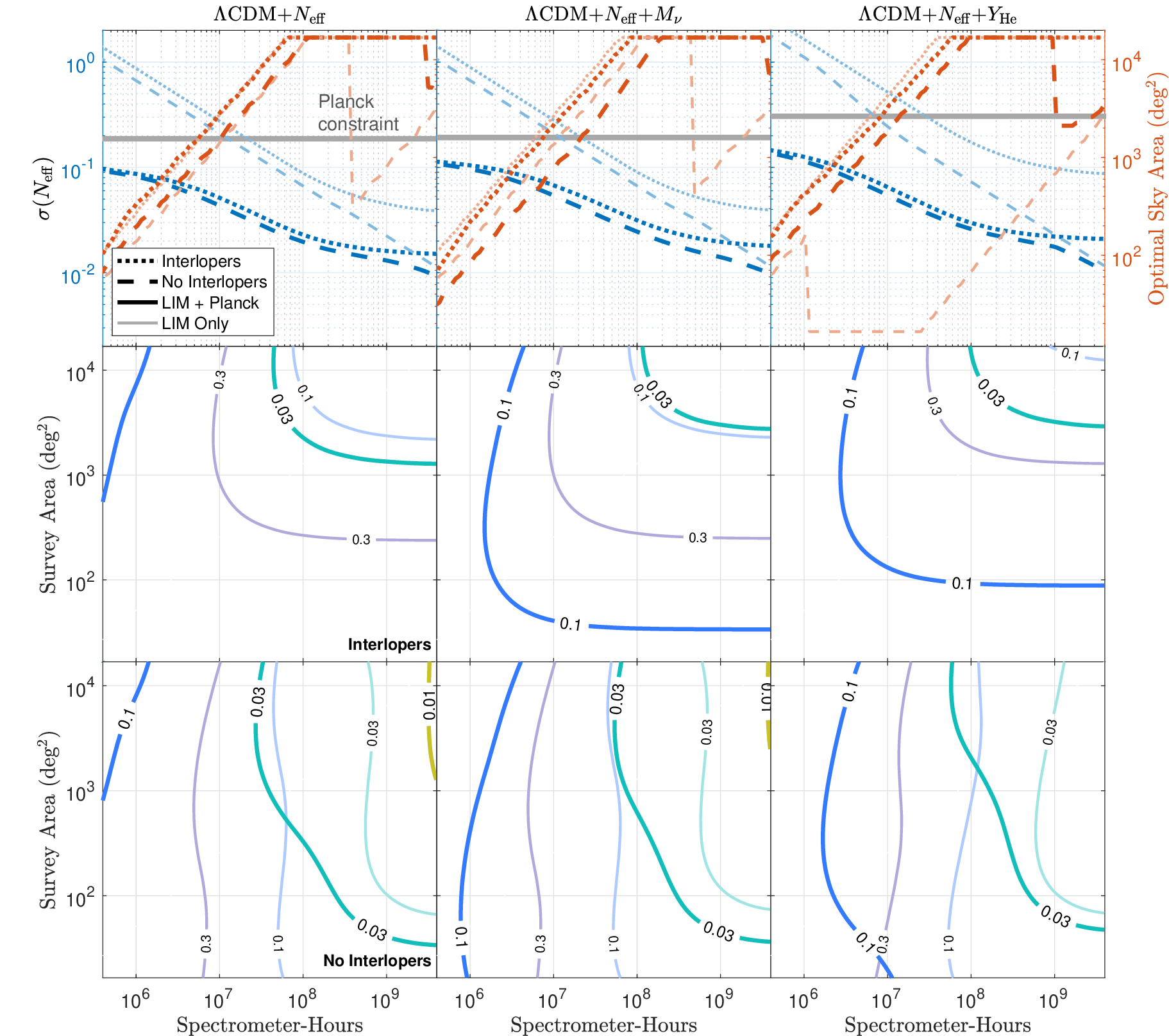}
    \caption{Optimized LIM surveys for probing light relics, from the combination of all six spectral lines. The columns from left to right correspond to varying only $N_{\rm eff}$, or together with $M_\nu$, or $Y_{\rm He}$. {\it Top row}: 1$\sigma$ marginalized constraints on $N_{\rm eff}$ as a function of spectrometer-hours (blue), and the corresponding sky coverage (red). The gray horizontal lines are Planck-only constraints. The lighter-shaded lines show the constraints from LIM-only, while the darker-shaded lines show the constraints from LIM+Planck. The dotted lines show the constraints with interlopers accounted for, and the dashed lines show the constraints when interloper emission is neglected. {\it Bottom two rows}: Contours of constant 1$\sigma$ errors on $N_{\rm eff}$ in the plane of survey area and spectrometer-hours, when interlopers accounted for (middle) and when interlopers are neglected (bottom). The colors of different contours are matched to specific values for $\sigma(N_{\rm eff})$ across all panels, to allow for ease of comparison. As is the case for the top row of panels, the lighter-shaded contours show the constraints from LIM-only, while the darker-shaded contours show the constraints from LIM+Planck.}
    \label{fig:optsurveys_neff}\vspace{.2in}
\end{figure*}

\begin{figure*}[t]
    \centering
    \flushleft\includegraphics[width=0.34\textwidth]{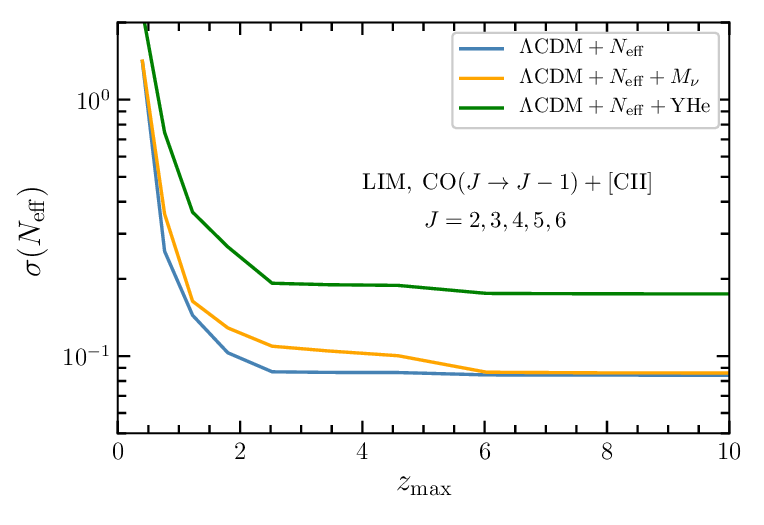}
    \hspace{-.13in} \includegraphics[width=0.34\textwidth]{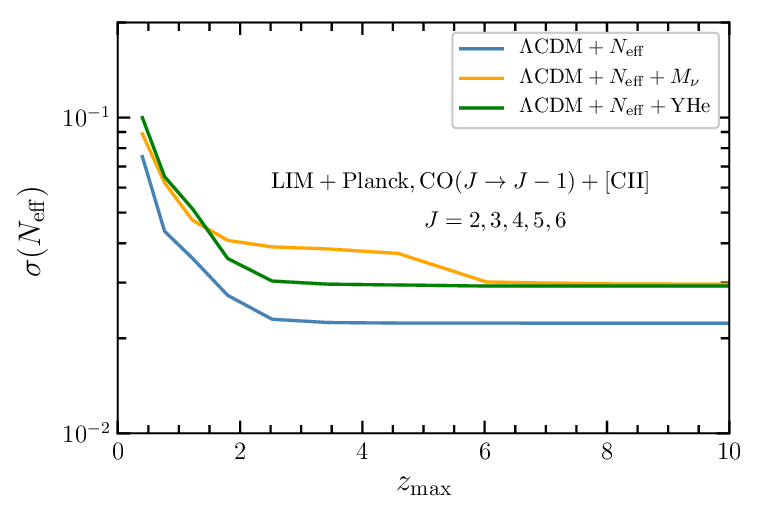}
    \hspace{-.13in }\includegraphics[width=0.34\textwidth]{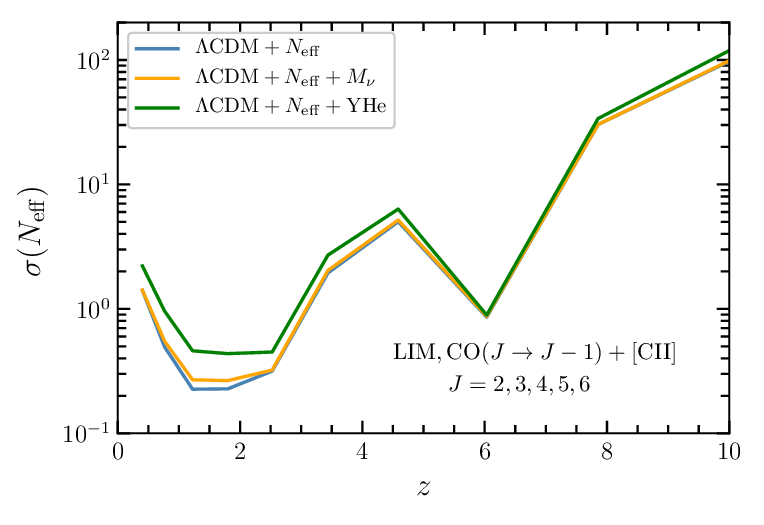}\vspace{-.1in}
    \caption{Redshift dependence of 1$\sigma$ marginalized constraints on $N_{\rm eff}$, using the combination of all six spectral lines. The first two panels on the left show the constraints for LIM-only (left) and Planck+LIM (middle) as a function of maximum redshift, $z_{\rm max}$, considered in the analysis (note the different vertical scales). The panel on the right shows the constraints per redshift bin, where $z$ is the median redshift of the bin. Different colors show the constraints when only varying $N_{\rm eff}$ (blue), or co-varying it with $M_{\nu}$ (orange), or with $Y_{\rm He}$ (green). An optimized survey with $1.45 \times 10^8$ spectrometer-hours is assumed, and interloper lines are accounted for.}
    \label{fig:sigz_neff}\vspace{.2in}
    
    \vspace{.1in}
    
    \centering
    \flushleft \begin{overpic}[scale=0.85]{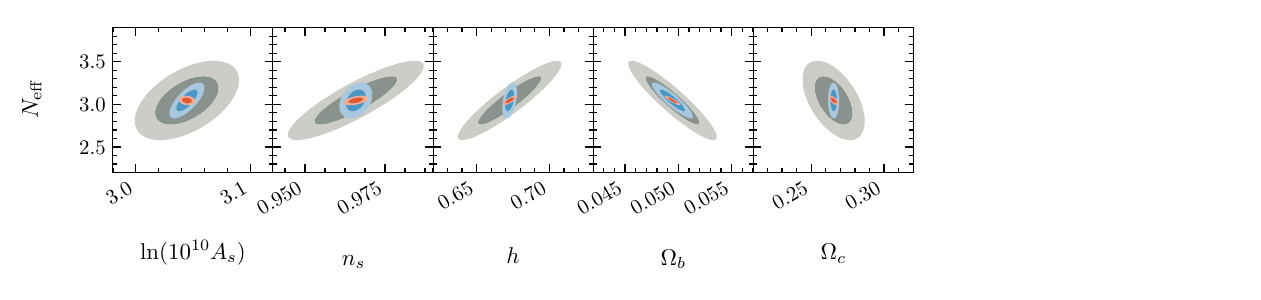}
     \put(80,12){\includegraphics[scale=0.85]{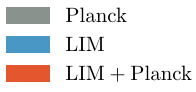}}  
  \end{overpic}
    \vspace{-.55in}
    \flushleft \includegraphics[scale=.85]{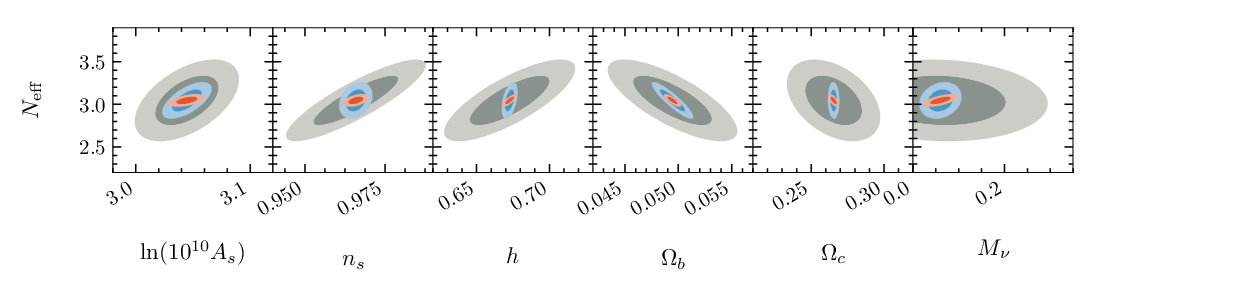}\vspace{-.4in}
    \flushleft\includegraphics[scale=.85]{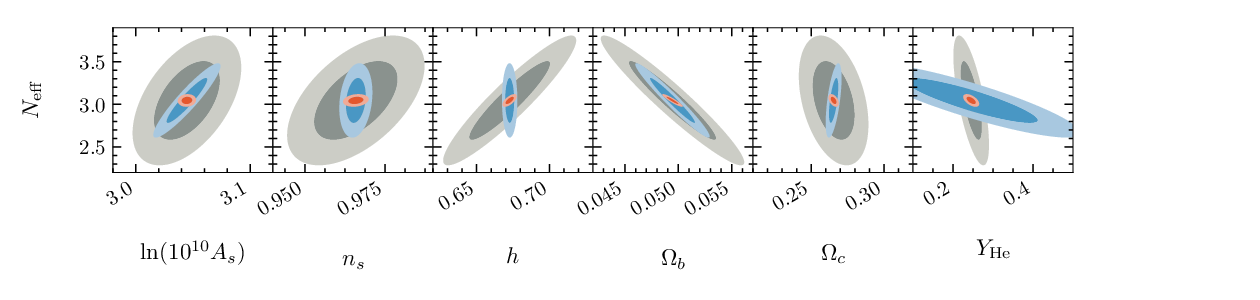}
    \vspace{-.3in}
    \caption{2D marginalized constraints on $N_{\rm eff}$ in 1- and 2-parameter extensions of $\Lambda$CDM from combination of all spectral lines considered in this paper. The rows from top to bottom correspond to varying only $N_{\rm eff}$ (top), varying $N_{\rm eff} + M_\nu$ (middle), and varying $N_{\rm eff} + Y_{\rm He}$ (bottom). Blue and gray contours correspond to constraints from Planck data and LIM individually, while the red contours are from combination of LIM and Planck. The LIM constraints correspond to an optimal survey with $1.45 \times 10^8$ spectrometer-hours, and interloper lines are accounted for.}
    \label{fig:neff_contours}
\end{figure*}

\subsection{Effective Number of Light Relics}
In the top row of Fig.~\ref{fig:optsurveys_neff}, as a function of spectrometer-hours, we show the best achievable 1$\sigma$ constraint on $N_{\rm eff}$ (in blue) and the optimal sky area for that constraint (in red). In the two bottom rows, we show the constant 1$\sigma$ contours in the plane of survey area vs.\ spectrometer-hours, accounting for interlopers (middle), and neglecting them (bottom). The columns show the constraints when varying only $N_{\rm eff}$ (left), varying it simultaneously with $M_\nu$ (middle), or with $Y_{\rm He}$ (right). In the top rows, the gray horizontal lines are Planck-only constraints, with the lighter lines correspond to LIM-only constraints, while darker lines are obtained from the combination of Planck and LIM. The dotted lines show the constraints with the effects of interloper emission included, while the dashed lines show the constraints with these effects neglected. The main observations can be summarized as follows: 
\begin{itemize}[leftmargin=.22in,itemsep=.02in]
    \item Considering LIM alone and neglecting the interlopers (light dashed lines in the top row), for all three cosmologies, $\sigma(N_{\rm eff})$ decreases as a power-law when increasing the number of spectrometer-hours, $\tau_{\rm sh}$. Simultaneous variation of $M_\nu$ has negligible effect, while variation of $Y_{\rm He}$ steepens the power-law, degrading the constraints at the lowest value of $\tau_{\rm sh}$ by about a factor of 2, and leaving the high $\tau_{\rm sh}$ end unaffected. Accounting for interlopers (light dotted blue lines) degrades the constraint, most notably when simultaneously varying $Y_{\rm He}$; for low spectrometer-hours, the scaling of $\sigma(N_{\rm eff})$ and $\tau_{\rm sh}$ stays the same as no-interloper case. However, the constraints begin to plateau when reaching $\tau_{\rm sh} \sim {\rm few} \times10^8$. This is because increasing spectrometer-hours cannot compensate for the additional noise from interlopers. While constraints on $N_{\rm eff}$ are nearly unaffected by simultaneous variation of $M_\nu$, varying $Y_{\rm He}$ degrades the constraints for both low and high values  of $\tau_{\rm sh}$ (about a factor of 2 for the latter). 
    \item When combining Planck and LIM (shown in dark dark blue lines in the top row), for low spectrometer-hours the constraints are dominated by Planck. Nevertheless, the addition of LIM improves the Planck constraints by about a factor of 2 at the lowest value of $\tau_{\rm sh}$. Similar to LIM-only case, when accounting for interlopers, the constraints reach a plateau as we increase $\tau_{\rm sh}$. It is worth noting that in combination with Planck, the impact of interlopers---in particular at high spectrometer-hours---is less significant compared to LIM-only constraints. When neglecting interlopers, enlarging the parameter space to vary $M_\nu$ or $Y_{\rm He}$ has negligible effect, while it degrades the constraints by about 30\% for both cases, when interlopers are accounted for.
    \item When considering LIM alone and neglecting interlopers, there exist multiple minima, as seen in the bottom row of panels. This arises from the fact that different line species, arising from separate redshifts with different brightnesses, require different survey depths (and correspondingly smaller $f_{\rm sky}$ for fixed spectrometer-hours) in order to achieve an optimal result. This apparent degeneracy in survey optimization is broken when moving to higher spectrometer-hours, as cosmic variance quickly becomes the dominant source of uncertainty for the low-redshift line species, and pushes the optimal design to deeper surveys (and correspondingly smaller sky areas) in order to probe the larger, higher-redshift cosmological volumes. This is the source of the large drop in optimal sky area seen in the top row of panels: as $\tau_{\rm sh}$ increases, a survey focused on higher-redshift emission overtakes a lower-redshift survey as the optimal choice.     
    \item As demonstrated in the bottom two rows of panels in Figure~\ref{fig:optsurveys_neff}, the constraining power of the survey is somewhat insensitive to the choice of $f_{\rm sky}$, with a reasonably broad range (better than an order of magnitude in several cases) of near-optimal choices provided that you are above the threshold at which cosmic variance strongly dominates the measurement errors, particularly in the case where interlopers are included. This appears to be a consequence of the broad redshift coverage of the proposed survey, where competing effects of cosmic variance and instrument noise for low and high-redshift, respectively, are relatively balanced in their impact on the constraining power.
\end{itemize}

We illustrate the information content of LIM at different redshifts in Fig.~\ref{fig:sigz_neff}, using an example survey with $1.45\times10^8$ spectrometer-hours\footnote{This value of spectrometer-hours corresponds roughly to a survey using a filled focal plane on an SPT-like instrument, which could be fielded in the next decade.}. The contribution of interloper lines to the noise (Eq.~\ref{eq:var}) is accounted for here, so these results are the worst-case interloper scenario. We show the 1$\sigma$ marginalized constraints on $N_{\rm eff}$ as a function of maximum redshift included in the analysis, $z_{\rm max}$, for LIM-only (left panel) and for Planck+LIM (middle panel). In the right panel, we show the $\sigma(N_{\rm eff})$ per redshift bin from LIM-only, where $z$ is the median redshift of the bin. Different colors show the constraints for various $\Lambda$CDM extensions. In the left and middle panels, we see that when holding $M_\nu$ constant (blue and green lines), the constraints nearly reach a plateau at $z \sim 2.5$. The reason for this saturation is that the main advantage of larger $z_{\rm max}$ for constraining $N_{\rm eff}$ is the increased mode count from a larger survey volume. However, at higher redshifts, the accessible volume increases more slowly than at lower redshifts. Therefore, the corresponding increase in volume provides little additional improvement to $\sigma(N_{\rm eff})$. On the other hand, when varying neutrino mass, at $z<4.5$, the orange line plateaus more slowly: higher $z_{\rm max}$ not only provides more modes, but also probes the redshift-dependent imprint of massive neutrinos on the power spectrum. 

Furthermore, at $ 4.5<z<6$ there is a downward step feature, and $\sigma(N_{\rm eff})$ reduces by $\sim 15\%$ for LIM-only and $\sim 20\%$ for Planck+LIM. This feature is due to the fact that at $z \geq 6$, in addition to the two highest CO rotational lines (J = 5, 6), we also detect the [CII] signal. Among these three lines, the constraining power lies predominantly with [CII] since it is the brightest at those redshifts (see Figs.~\ref{fig:tbar}, \ref{fig:line_pk}), and the instrument considered in this work has the highest sensitivity to it. When considering the constraints per redshift bin (right panel of Fig.~\ref{fig:sigz_neff}), in all three cosmologies considered, $\sigma(N_{\rm eff})$ improves at $z\sim6$. However, in the cumulative constraints summed over all redshifts up to a $z_{\rm max}$, the step feature is only seen when neutrino mass is varied---here a longer redshift arm probes the redshift-dependence of the imprints of neutrino mass. Finally, from the first plot on the left we see that summed over all redshift bins, extending the parameter space to include variation of $Y_{\rm He}$ degrades the LIM-only constraint on $N_{\rm eff}$ by about a factor of 2; leaving the sum of neutrino masses as a free parameter does not have a significant effect. When combined with Planck data, the 2-parameter extensions degrade the $N_{\rm eff}$ constraint similarly at high $z_{\rm max}$, degrading by about a factor of 1.3 compared to the model with only $N_{\rm eff}$ free. 

To highlight the importance of alleviating parameter degeneracies when combining LIM and CMB data, for an optimized survey with $\tau_{\rm sh} =1.45\times10^8$, in Fig.~\ref{fig:neff_contours} we show the 2D marginalized constraints on $N_{\rm eff}$ vs.\ other cosmological parameters, with Planck data in gray, LIM in blue, and their combination in red. The rows correspond to varying only $N_{\rm eff}$ (top), or varying it together with $M_\nu$ (middle) or $Y_{\rm He}$ (bottom). The full triangle plots for the base model of $\Lambda$CDM+$N_{\rm eff}$ are shown in Appendix~\ref{app:EUCLID_Comp}. In CMB data, various degeneracies weaken the constraint on $N_{\rm eff}$. LIM alone provides significantly tighter constraints on all parameters, except for $Y_{\rm He}$, compared to Planck. Thanks to the difference in parameter degeneracy directions in CMB and LSS (most notably between $N_{\rm eff}$ and $n_s, h,\Omega_c, Y_{\rm He}$), when combining LIM with Planck, the parameter constraints improve further. The difference in degeneracy directions can be most clearly seen in the case of $h$ and $Y_{\rm He}$, and to a lesser extent for $n_s$ and $\Omega_c$.

\begin{figure*}[t]
    \centering
    \includegraphics[width=\textwidth]{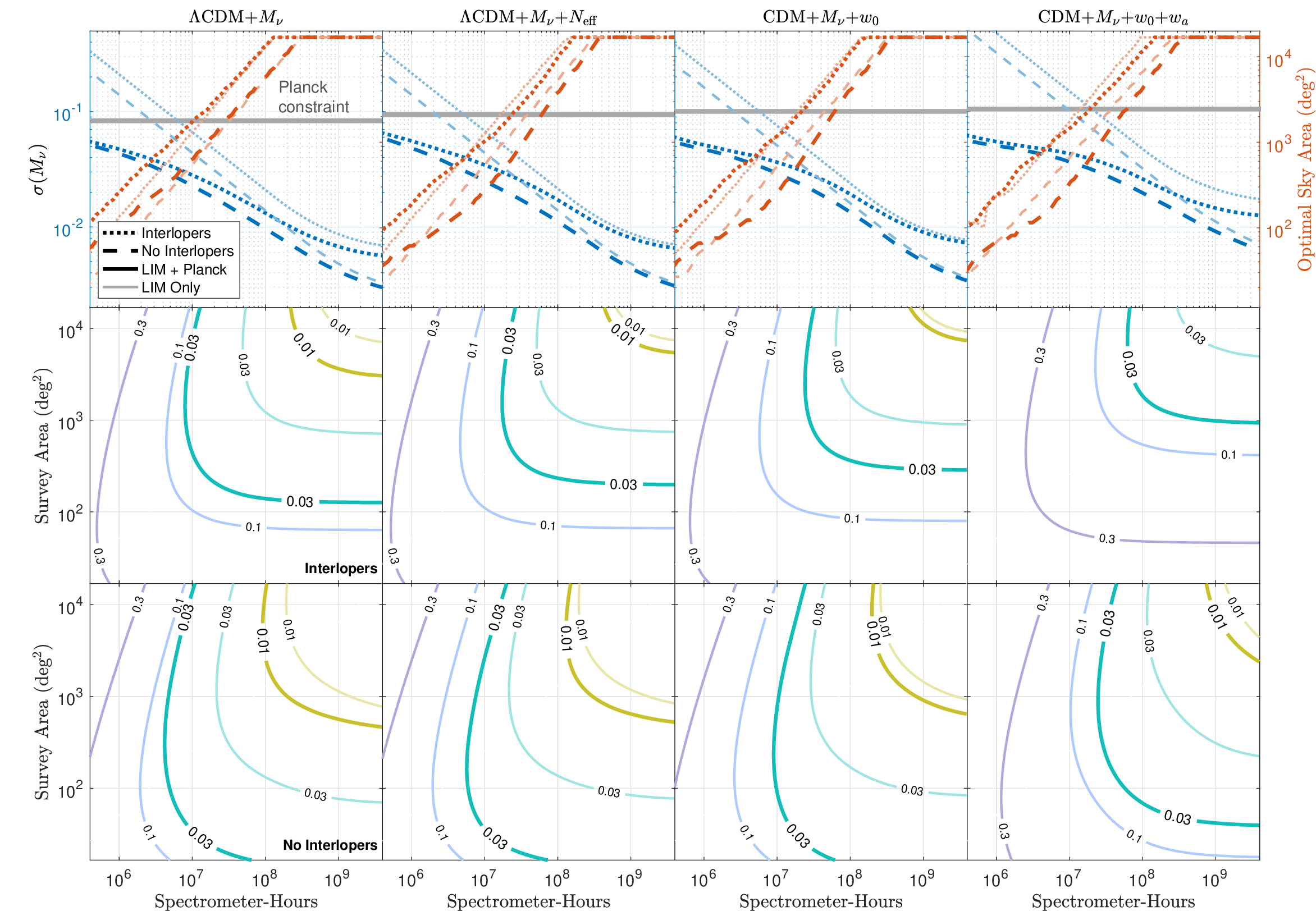}
    \caption{Optimized LIM surveys for probing the sum of neutrino masses, using the combination of all six spectral lines. The columns from left to right correspond to varying only $M_\nu$, varying it together with $N_{\rm eff}$, or $w_0$, or $w_0+w_a$. {\it Top row}: The blue lines show 1$\sigma$ marginalized constraints on $M_\nu$, as a function of spectrometer-hours. The red lines show the corresponding sky coverage. The gray horizontal lines are Planck-only constraints. The lighter-shaded lines show the constraints from LIM-only, while the darker-shaded lines show the constraints from LIM+Planck. The dotted lines show the constraints with interlopers accounted for, and the dashed lines show the constraints when interloper emission is neglected. {\it Bottom two rows}: Contours of constant 1$\sigma$ errors on $M_\nu$ in the plane of survey area and spectrometer-hours, when interlopers accounted for (middle) and when interlopers are neglected (bottom). The colors of different contours are matched to specific values for $\sigma(M_{\nu})$ across all panels, to allow for ease of comparison. As is the case for the top row of panels, the lighter-shaded contours show the constraints from LIM-only, while the darker-shaded contours show the constraints from LIM+Planck.}
    \label{fig:optsurveys_mnu}

    \vspace{.1in}
    
    \flushleft\includegraphics[width=0.34\textwidth]{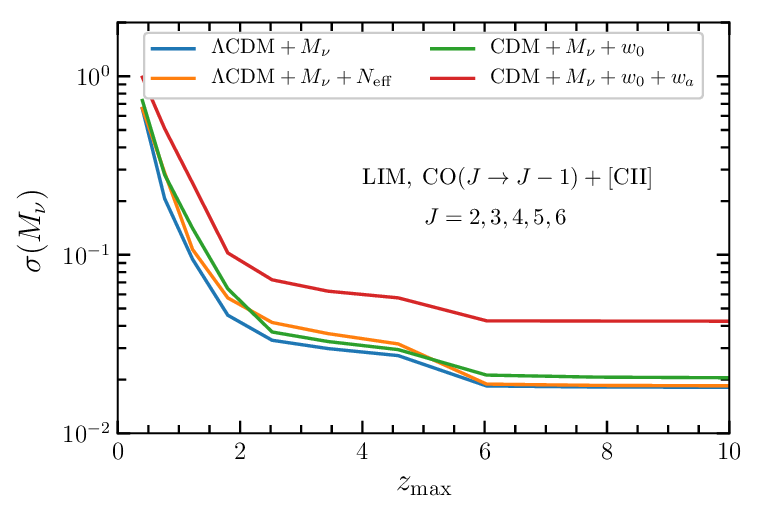}
    \hspace{-.13in}\includegraphics[width=0.34\textwidth]{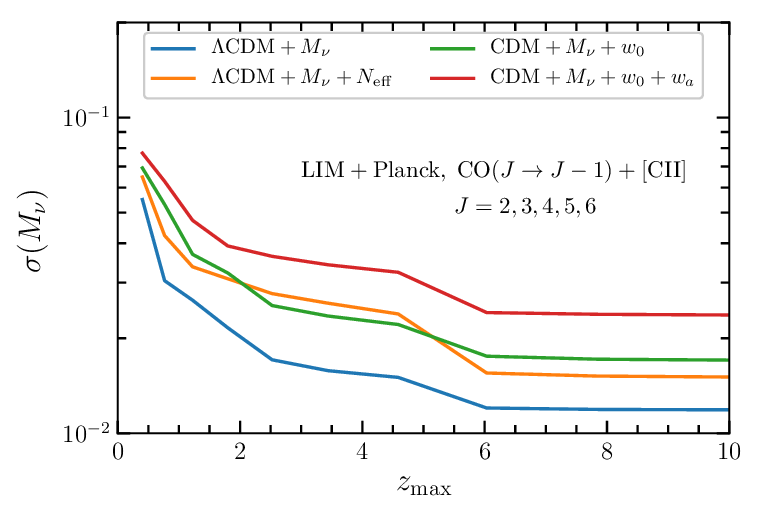}    
   \hspace{-.13in}\includegraphics[width=0.34\textwidth]{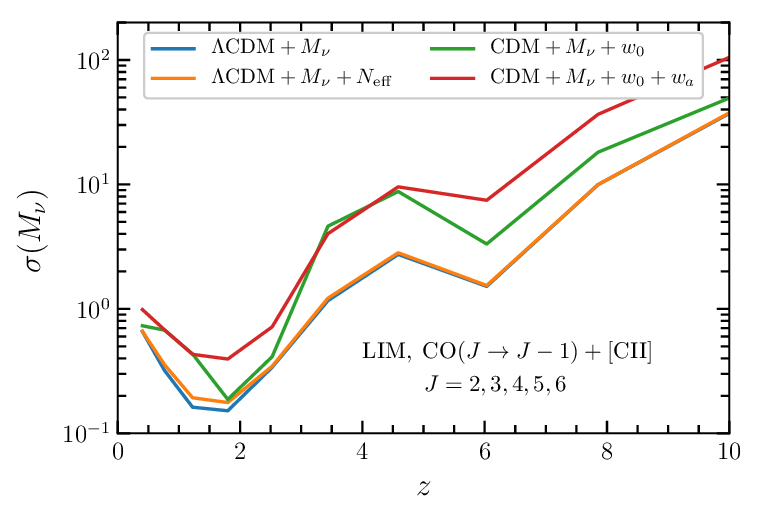}\vspace{-.1in}
   \caption{Redshift-dependence of 1$\sigma$ marginalized constraints on  $M_\nu$ (in unit of [${\rm eV}$]), from the combination of all six spectral lines. The first two plots on the left show the constraints for LIM-only (middle) and Planck+LIM (right) as a function of maximum redshift, $z_{\rm max}$, while the panel on the right shows the constraints per redshift bin, where $z$ is the median redshift of the bin. Different colors correspond to only varying $M_\nu$ (blue), or co-varying it with $N_{\rm eff}$ (orange), with $w_0$ (green), or with $w_0+w_a$ (red). An optimized survey with $1.45 \times 10^8$ spectrometer-hours is assumed, and interloper lines are accounted for.}
    \label{fig:sigz_mnu}
\end{figure*}
 
\begin{figure*}[t]
    \centering
     \flushleft \begin{overpic}[scale=0.85]{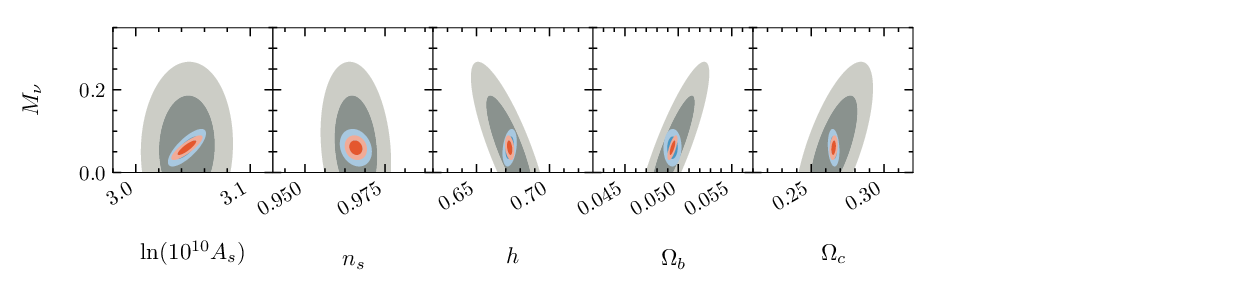}
     \put(80,12){\includegraphics[scale=0.85]{legend.pdf}}  
  \end{overpic}
    \vspace{-.55in}
     \flushleft\includegraphics[scale=.85]{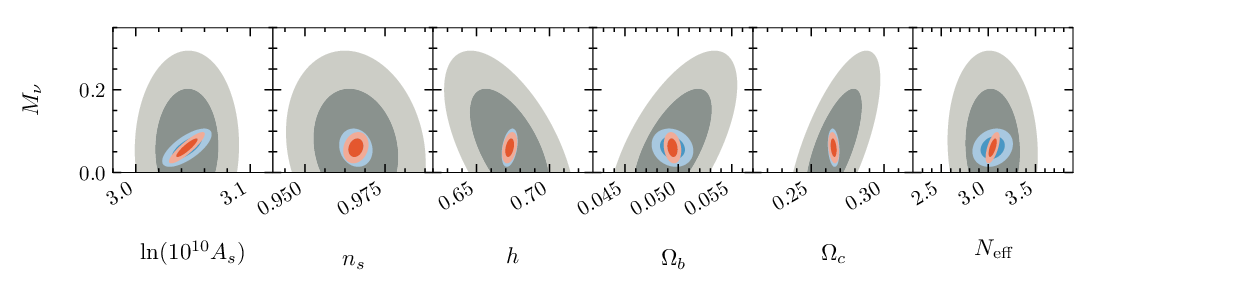}\vspace{-.4in}
     \flushleft\includegraphics[scale=.85]{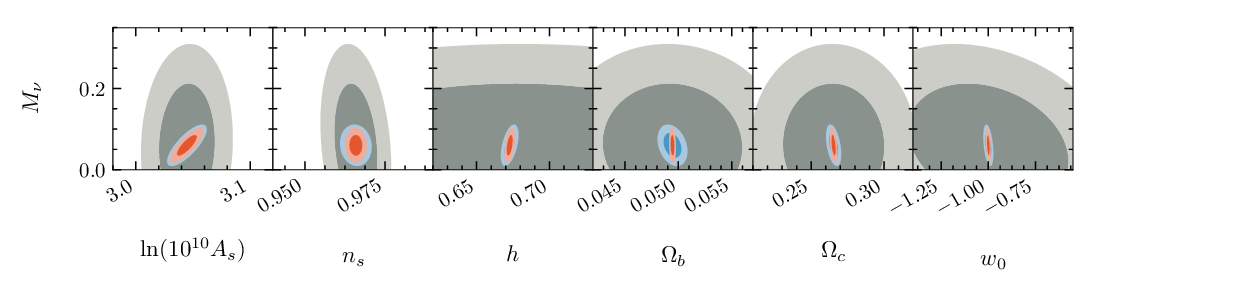}\vspace{-.4in}
    \flushleft\includegraphics[scale=.85]{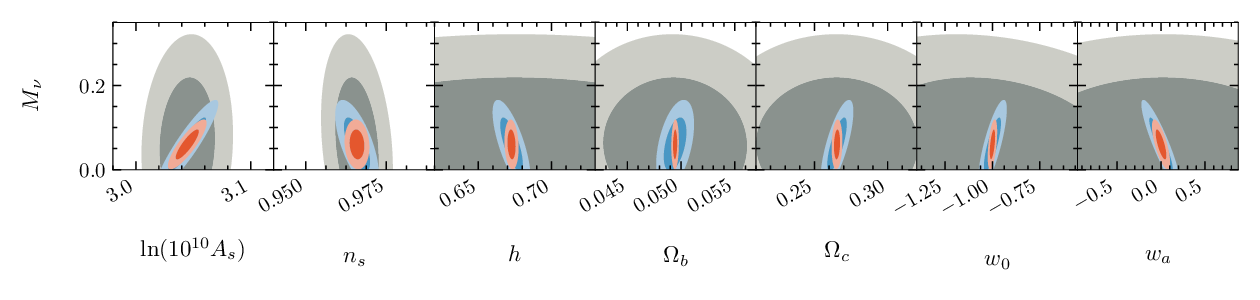}
    \vspace{-.3in}
    \caption{2D marginalized constraints on $M_\nu$ from combination of all of the spectral lines considered in this paper. Rows from top to bottom correspond to varying only $M_\nu$ (first row), varying it together with $N_{\rm eff}$ (second row), with $w_0$ (third row), or with $w_0+w_a$ (fourth row). Color coding and LIM specifications are the same as Fig. \ref{fig:neff_contours}.}    
    \label{fig:mnu_contours} \vspace{.15in}
\end{figure*}

\subsection{Sum of Neutrino Masses}

In the top row of Fig.~\ref{fig:optsurveys_mnu}, as a function of spectrometer-hours, we show the best achievable 1$\sigma$ constraint on $M_\nu$ and the optimal sky area for that constraint. The bottom panels show the constant $1\sigma$ contours in the plane of survey area and spectrometer-hours. The columns from left to right show the constraints when varying only $M_\nu$ (first), varying it simultaneously with $N_{\rm eff}$ (second), $w_0$ (third) and $w_0+w_a$ (fourth). The line styles and colors are the same as in Fig.~\ref{fig:optsurveys_neff}. The main observations from these plots can be summarized as follows: 
\begin{itemize}[leftmargin=.22in,itemsep=.02in]
    \item The top row shows that when considering LIM alone and neglecting interlopers (dashed light blue lines), for all models considered, $\sigma(M_\nu)$ decreases nearly as a power-law when increasing spectrometer-hours, $\tau_{\rm sh}$ (the curves very slowly flatten at the highest end). Simultaneous variation of $N_{\rm eff}$ does not degrade the constraints. On the other hand, assuming dynamic DE and varying both $w_0$ and $w_a$ increases $\sigma(M_\nu)$ by a factor of $\sim 2.5$; assuming a constant EoS and varying only $w_0$ degrades the constraint only mildly (at most by $\sim 25\%$). Including interlopers as a source of noise (dotted light blue lines) degrades the constraint with a similar trend in all four cosmologies considered. For low values of spectrometer-hours, the scaling of $\sigma(M_\nu) - \tau_{\rm sh}$ stays the same as in the no-interloper case. However, the constraints approach a plateau at $\tau_{\rm sh} \sim 10^9$. As in the case of $N_{\rm eff}$ discussed previously, this is due to the interlopers starting to dominate the error budget. 
    \item The top row further shows that for the combination of Planck+LIM with and without interlopers (dotted and dashed dark blue lines), constraints are dominated by Planck data at low values of $\tau_{\rm sh}$. Nevertheless, the addition of LIM improves the Planck constraints by $\sim30\%$ when varying $M_\nu$ alone or together with $N_{\rm eff}$, and $\sim40\%$ when also varying the dark energy parameter. Including interlopers, the constraints approach a plateau as we increase $\tau_{\rm sh}$. Different from the case of $N_{\rm eff}$, at the high end of spectrometer-hours, the LIM-only and Planck+LIM curves trace each other closely since LIM almost entirely dominates the constraining power.
    \item Generally, we find that the optimal $f_{\rm sky}$ increases roughly as $\tau_{\rm sh}^{2/3}$, which can be understood as maintaining the balance between uncertainties driven by cosmic variance (which roughly decreases as $f_{\rm sky}^{-0.5}$) and instrument noise (which at fixed volume, decreases inversely with $\tau_{\rm sh}$). At the low end of spectrometer-hours ($\lesssim10^{6}$) for Planck+LIM, the slope is shallower, due to Planck priors that provide the bulk of the constraining power for $\sigma(M_{\nu})$. The primary additive strength is in breaking the degeneracies in Planck data.
    \item Similar to the case with $N_{\rm eff}$, the constraining power of a given survey is somewhat insensitive to the choice of $f_{\rm sky}$, provided that you are above the threshold at which cosmic variance strongly dominates the measurement errors.
\end{itemize}

In the left and middle panels of Fig.~\ref{fig:sigz_mnu}, we plot $\sigma(M_\nu)$ as a function of $z_{\rm max}$ from LIM alone and from its combination with Planck. In the right panel, we show $\sigma(M_\nu)$ per redshift bin, with z being the median redshift of the bin. The survey specifications are the same as in Fig.~\ref{fig:sigz_neff}, and constraints for various extensions of $\Lambda$CDM are shown in different colors.  When considering LIM alone, enlarging the parameter space does not degrade the constraints dramatically, except for when assuming redshift-dependent dark energy and varying both $w_0$ and $w_a$. For the combined LIM and Planck data, however, varying additional parameters increases $\sigma(M_\nu)$ more considerably. Again, we see the same downward step features as in Fig.~\ref{fig:sigz_neff} at $ 4.5 < z < 6$ as a result of the additional [CII] signal at $z \geq 6$. For $\Lambda$CDM+$M_\nu$ cosmology, the $1\sigma$ errors on $M_\nu$ reduce by $\sim30\%$ for LIM alone and by $\sim20\%$ for Planck+LIM. The plateau at $z>6$ is due to larger instrument and interloper noise. 

We note that at $z < 4.5$, compared to $\sigma(N_{\rm eff})$ in Fig.~\ref{fig:sigz_neff}, the constraints approach a plateau more slowly. As discussed before, going to higher $z_{\rm max}$ not only provides more modes, but also allows for probing the redshift-dependent imprint of massive neutrinos on the power spectrum. The per-bin constraints shown in the right panel of Fig.~\ref{fig:sigz_mnu} indicate that the improvement at $z\sim 6$ (seen in the cumulative plots on the right) is driven not only by additional [CII] signal at $z\sim 6$, but also by having a long redshift arm to probe the redshift evolution of the suppression of the matter power spectrum and growth rate of structure.

Finally, in Fig.~\ref{fig:mnu_contours}, we show the 2D marginalized constraints on $M_\nu$ for an optimized survey with $1.45\times10^8$ spectrometer-hours, for Planck (in gray), LIM (in blue), and their combination (in red). The rows from top to bottom correspond to varying only $M_\nu$ (first), or co-varying it with $N_{\rm eff}$ (second), $w_0$ (third) and $w_0+w_a$ (fourth). The full triangle plots for the base model of $\Lambda$CDM+$M_\nu$ are shown in Appendix~\ref{app:EUCLID_Comp}. Again, LIM data by itself significantly improves the constraints on all cosmological parameters. Apart from neutrino mass and $\Lambda$CDM parameters, LIM data provide very tight constraints on the dark energy equation of state (see Table~\ref{tab:1sig} for $1\sigma$ constraints). Additionally, the combination with Planck tightens constraints further by breaking parameter degeneracies thanks to different degeneracy directions for CMB and LSS observables. Enlarging the parameter space affects the LIM constraints much less significantly than CMB (temperature+polarization) data. This is most notable when including variation of dark energy, for which CMB primary anisotropies largely lose their constraining power.  \vspace{-.05in} \\

\begin{deluxetable*}{ccc|cc|cc}[h]
\tablehead{
\colhead{Spec-hrs} & \colhead{Example} & \colhead{Deployment Timescale} & \multicolumn{2}{c}{$\sigma(M_{\nu})$ [eV]} & \multicolumn{2}{c}{$\sigma(N_{\rm eff})$} \\
\colhead{ } & \colhead{ } & \colhead{ } &
\colhead{Int.} & \colhead{No Int.} &
\colhead{Int.} & \colhead{No Int.}
}
\tablecaption{Future Survey Capabilities\label{tab:stages}}
\startdata
$10^5$ & TIME              &  Now  & 0.69 (0.066)    & 0.48 (0.061)    & 2.8 (0.11)    & 2.0 (0.10)   \\ \hline
$10^6$ & TIME-EXT          &  3 yr & 0.21 (0.047)    & 0.14 (0.043)    & 0.87 (0.087)  & 0.67 (0.082) \\ \hline
$10^7$ & SPT-like, 1 tube  &  4 yr & 0.066 (0.028)   & 0.044 (0.023)   & 0.27 (0.051)  & 0.21 (0.043) \\ \hline
$10^8$ & SPT-like, 7 tubes &  8 yr & 0.021 (0.013)   & 0.014 (0.0097)  & 0.088 (0.023) & 0.0674 (0.020)\\ \hline
$10^9$ & CMB-S4-like, 85 tubes & 12 yr & 0.0087 (0.0068) & 0.0048 (0.0041) & 0.045 (0.016) & 0.022 (0.013)
\enddata
\tablecomments{Potential stages of future mm-wave LIM experiments and corresponding neutrino constraints. Values provided are for LIM-only (with Planck+LIM in parentheses), for the best- and worst-case interloper scenarios. For each stage of experiment, we provide an approximate example of the class of instrument required for such a survey. For future instruments, an optics tube is assumed to hold a focal plane of $\sim 400$ on-chip spectrometers observing from 80--310 GHz. Future surveys are assumed to run for multiple years, observing for several thousand hours per year. The timescale is a rough estimate to when such a survey could begin operations. } \vspace{.1in}
\end{deluxetable*}

\section{Conclusions}\label{sec:conclusion}

In this work we have forecasted the constraining power of next-generation ground-based mm-wave LIM experiments on $N_{\rm eff}$ and $M_{\nu}$. Over a wide range of experimental sensitivities, for a variety of extensions to $\Lambda$CDM we used the Fisher formalism to determine the tightest possible constraints and the optimal survey area, evaluating both the best- and worst-case interloper line mitigation scenarios. We considered a range of  experimental sensitivities, as summarized in Table~\ref{tab:stages}, along with a rough estimate for the time at which such surveys could start, driven by the anticipated increase in density of on-chip spectrometers.

One of the primary advantages of mm-wave LIM is cost-effectiveness, especially compared to contemporary optical galaxy surveys and space missions. While the detector technology is still being developed, mm-wave spectrometers draw from a long heritage of CMB experience in mass-producing densely-packed, background-limited detectors. Completing the R\&D for compact mm-wave spectrometers and outfitting an existing instrument such as the SPT with a full complement of $R=300$ spectrometers would enable a LIM survey of the order $\sim 10^8$ spectrometer-hours to be completed. Such a survey could be deployed by the end of the decade, at significantly lower cost than experiments such as CMB-S4 or Euclid.

We have shown that with a conservative assumption of no removal of line interloper noise, an optimized survey of $\sim 10^8$ spectrometer-hours covering 40\% of the sky, combined with Planck, can constrain the effective number of light relics at the level of  $\sigma(N_{\rm eff}) \simeq 0.023 $, providing a $1.2\sigma$ exclusion of the minimal thermal abundance. For the sum of neutrino masses, such a survey would reach the precision of $\sigma(M_\nu) \simeq 13 \ {\rm meV}$, providing $\sim5\sigma$ (8$\sigma$) detection of the minimum neutrino mass in the normal (inverted) hierarchy. In comparison to constraints from CMB-S4 and Euclid---the latter shown in further detail in Appendix~\ref{app:EUCLID_Comp}---such a survey would provide meaningful and complementary contributions to constraints on $M_{\nu}$ and $N_{\rm eff}$.

We show results both for LIM alone and combined with Planck to illustrate parameter degeneracies that LIM helps to alleviate. In particular, we note that even more modest surveys of $10^{6}$--$10^{7}$ spectrometer-hours can significantly improve constraints on $N_{\rm eff}$ and $M_{\nu}$ by breaking degeneracies in Planck data. As discussed earlier, more realistic modeling will affect our constraints on cosmological parameters. This will include extending the linear model of the line power spectrum to include one-loop corrections, and marginalizing over additional line biases introduced at the one-loop level \citep{Sailer:2021yzm}. We leave quantification of these effects to future work. 

While we have only considered the information content of the line power spectrum in combination with Planck, exploiting higher-order statistics and synergies between LIM and future CMB and galaxy surveys would not only improve the forecasted constraints, but also offer a means to overcome degeneracies with nuisance astrophysical parameters. Furthermore, such information could improve the mitigation of systematics and foregrounds. We leave further studies in these directions to future work. 

While our focus in this paper has been on constraining neutrino properties, the discussed LIM surveys offer unique opportunities for a multitude of other science goals. Our results show that the considered surveys can provide exquisite constraints on $\Lambda$CDM parameters and the dark energy equation of state (see Table~\ref{tab:1sig}). The combination of Planck+LIM would achieve $\sigma(w_0) = 0.0051 $ and $\{ \sigma(w_0) \simeq 0.0098, \ \sigma(w_a) \simeq 0.041 \}$, assuming $\sim 10^8$ spectrometer-hours. The wide redshift range probed by such surveys would uniquely constrain the redshift-dependence of the expansion history, dark energy models, and modifications to gravity (see e.g., \citealt{Karkare:2018sar,Bernal:2019gfq} for existing forecasts for non-21cm lines and \citealt{Lorenz:2017fgo,Sailer:2021yzm} for 21cm). Furthermore, the large sky coverage and wide redshift range of such surveys make them particularly well-suited to probe primordial non-Gaussianity (see e.g., \citealt{MoradinezhadDizgah:2018zrs,MoradinezhadDizgah:2018lac,Liu:2020izx,Viljoen:2021ypp}). This wide range of potential science targets and unique opportunity to probe fundamental physics are strong motivations for developing the instrumental and observational techniques needed for high-sensitivity mm-wave LIM surveys.

An important caveat to the above is that while quite mature for CMB and galaxy surveys, instrument hardware and analysis methods for LIM surveys are still in a relatively nascent state, with the field primarily focused on pathfinder instruments and initial detections. Millimeter-wave LIM instruments have an advantage of a long heritage of CMB experiments, focused on similar wavelengths with well-developed sites and observation strategies, which in addition to reducing costs can also help mitigate the impact of systematics---a significant challenge for observations at longer wavelengths (e.g., 21cm cosmology; \citealt{Nasirudin_2020}). While pathfinder experiments will help to pave the path forward, the success of such large-scale surveys will require continued technical development, both in instrumentation and analysis tools.  

Should such efforts prove successful, surveys even larger than our nominal $10^{8}$ spectrometer-hours would have significant additive value. At the maximum survey scale considered here ($4\times10^{9}$ spectrometer-hours), in combination with Planck, one could constrain $\sigma(N_{\rm eff}) \simeq 0.015$ and $\sigma(M_\nu) \simeq 5.6 \ {\rm meV}$. While this would be a significant undertaking, these improvements suggest that such a survey is worth further consideration. At such large numbers of spectrometer-hours, the constraints become primarily limited by cosmic variance, so that sites that can observe a larger sky fraction become highly desirable. A space-based mission would be capable of measuring the largest possible sky area, and our analysis suggests that constraints on neutrinos---along with other cosmological parameters---may be strong science motivators for potential futuristic space-based LIM surveys now being discussed (e.g., \citealt{Delabrouille:2019thj,Silva2021}). 

\begin{acknowledgments}
It is our pleasure to thank Emanuele Castorina for many insightful discussions as well as his feedback on the draft. We also thank Amol Upadhye, Enea Di Dio, Steen Hannestad for helpful discussions, and Adam Anderson and Clarence Chang for their detailed comments on the draft of this paper. Finally, we thank Ana Diaz for collaborations at the very early stages of this work. A.M.D. is supported by the SNSF project ``The  Non-Gaussian  Universe and  Cosmological Symmetries", project number:200020-178787. A.M.D also acknowledges partial support from Tomalla Foundation for Gravity. 
\end{acknowledgments}

\vspace{.2in}
\appendix

\section{Redshift binning and instrument noise} \label{app:binning}
In Table~\ref{tab:noise_per_redshift_bin} we show the value of $P_{\rm N}$ used for each redshift bin, provided the instrument and atmospheric parameters in Section~\ref{sec:design}, for our smallest survey area (16.5 ${\rm deg}^2$) and minimum integration time ($2\times10^{5}$ spectrometer-hours). For surveys of different times and integration times, we use Eq.~\ref{eqn:noisepermode} to estimate the noise, which dictates that $P_{\rm N}\propto\Omega_{s}\tau_{sh}^{-1}$.   \vspace{.1in}

\section{Comparison with Euclid}\label{app:EUCLID_Comp}

In this appendix, we show the forecasted parameter constraints for the models described in Table~\ref{tab:models} from the Euclid spectroscopic sample combined with Planck temperature and polarization data. We also show the results from LIM-only and its combination with Planck. For Euclid, we use the specifications and given in Table 3 of the recent official Euclid forecast paper \citep{Blanchard:2019oqi}. The full survey covers an area of 15000 ${\rm deg}^2$, observing H$\alpha$ emitters in the redshift range of $0.9<z<1.8$, and binning the data in four redshift bins. We refer to the aforementioned paper for the expected values of shot noise and linear biases.

Analogous to line intensity power spectrum, we model the galaxy power spectrum assuming linear perturbation theory, and include the RSD and AP effect. Therefore, we have
\be
    P_g(k,\mu,z) = \frac{H_{\rm true}(z)}{H_{\rm ref}(z)}  \left[\frac{D_{A,{\rm ref}}(z)}{D_{A,{\rm true}}(z)}\right]^2 {\rm exp}\left(-\frac{k_{\rm true}^2 \mu_{\rm true}^2 \sigma_v^2}{H^2(z)}\right)  \left[1+\mu_{\rm true}^2\beta(k_{\rm true},z)\right]^2 b_g^2(z) P_0(k_{\rm true},z).
\ee
For spectroscopic galaxy sample, the $\sigma_z$ in Eq.~\eqref{eq:disp} represents the spectroscopic redshift error, which for Euclid is given by $\sigma_z = 0.001 (1+z)$. For each redshift bin, we set the value of $k_{\rm max}$ as described in Section \ref{sec:Fisher}.

\begin{figure*}[htbp!]
    \centering
    \includegraphics[width=.65\textwidth]{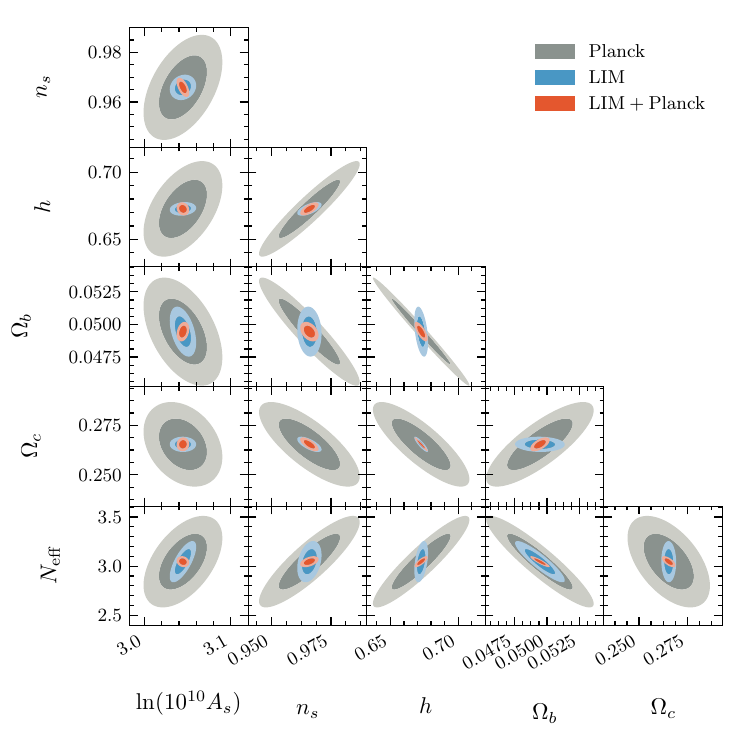}\vspace{-.15in}
    \includegraphics[width=.65\textwidth]{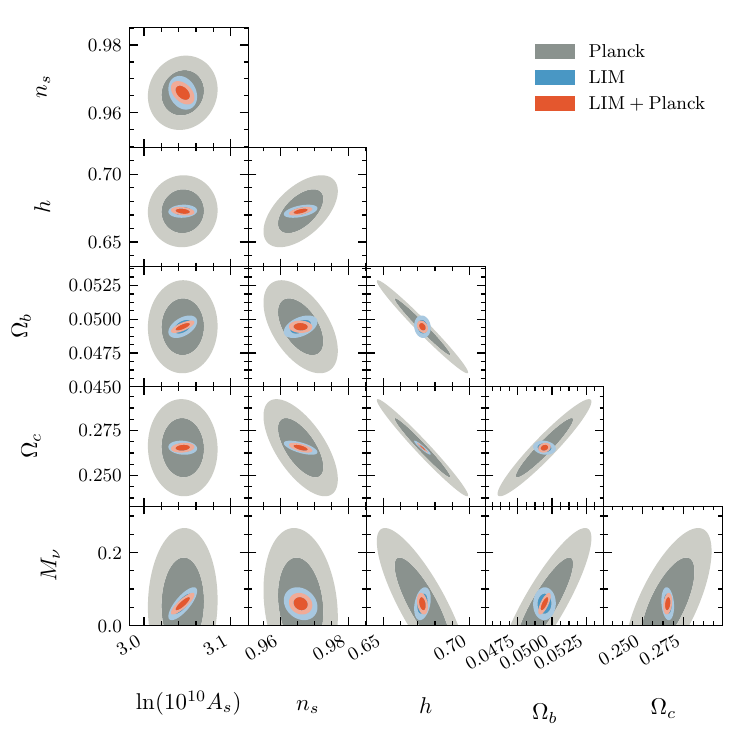}\vspace{-.2in}
     \caption{2D marginalized constraints on cosmological parameters for $\Lambda{\rm CDM}+N_{\rm eff}$ (top) and $\Lambda{\rm CDM}+M_\nu$ model (bottom), from Planck and LIM data alone and combined. LIM constraints correspond to an optimal survey with $10^8$ spectrometer-hours, and interloper lines are accounted for.}
    \label{fig:LIM_triangle}
\end{figure*}

In Table~\ref{tab:1sig} we show the 1$\sigma$ constraints from Planck data, Planck+Euclid and Planck+LIM. In the baseline cosmologies (1-parameter extensions to $\Lambda$CDM), our constraints from the combination of Euclid and Planck are in broad agreement with values reported in \cite{Obuljen:2017jiy} and \cite{Sprenger:2018tdb}. To demonstrate the parameter degeneracies, in Fig.~\ref{fig:LIM_triangle} we show the 2D marginalized errors for the full set of cosmological parameters when considering Planck (gray), LIM (blue) data alone and combined (red), while in Fig.~\ref{fig:LIM_triangle} we show the corresponding plots for Euclid. In each figure, the top plot shows the constraints for $\Lambda$CDM + $N_{\rm eff}$, while the bottom is for $\Lambda$CDM + $M_\nu$. Note that the combination of galaxy lensing and clustering measurements from Euclid will provide tighter constraints on neutrino properties compared to those reported here using spectroscopic clustering only (see e.g., \citealt{Sprenger:2018tdb}).

\begin{deluxetable*}{c|cccccccccc}[t]
\tablehead{\colhead{Line} & \multicolumn{10}{c}{Median Redshift ($z$)} \\
\colhead{Species} & \colhead{0.40} & \colhead{0.77} & \colhead{1.2} & \colhead{1.8} & \colhead{2.5} & \colhead{3.4} & \colhead{4.6} & \colhead{6.0} & \colhead{7.9} & \colhead{10.1}}
\tablecaption{Estimated per-mode instrument noise \label{tab:noise_per_redshift_bin}}
\startdata
    CO(2-1)   & 2.8  & 3.2  & 3.7 & 3.7 & --- & --- & --- & --- & --- & --- \\
    CO(3-2)   & 2.4  & 3.6  & 3.2 & 3.8 & 4.2 & --- & --- & --- & --- & --- \\
    CO(4-3)   & ---  & 3.0  & 3.3 & 3.6 & 3.7 & 4.1 & 4.4 & --- & --- & --- \\
    CO(5-4)   & ---  & ---  & 3.4 & 3.5 & 3.8 & 3.7 & 4.1 & 4.5 & --- & --- \\
    CO(6-5)   & ---  & ---  & --- & 3.3 & 4.2 & 3.6 & 4.1 & 4.3 & --- & --- \\
    $[$CII$]$ & ---  & ---  & --- & --- & --- & --- & --- & 3.7 & 3.9 & 3.8 \\
\enddata 
\tablecomments{Values are given in units of $\log_{10}[\mu {\rm K}^{2} \, (h/\textrm{Mpc})^{3}]$, covering 16.5 ${\rm deg}^2$, with $4\times10^{4}$ spectrometer-hours, following the instrument parameters and weather conditions provided in Section~\ref{sec:design}. Redshift bins are spaced such that they are 0.1 dex wide, to account for instrument noise variation and redshift evolution of the line-emitting population of sources.} 
\end{deluxetable*}

\begin{table*}[t] \vspace{-2.5in}
\caption{1$\sigma$ marginalized constraints on cosmological parameters from Planck data alone, Planck combined with Euclid (galaxy clustering), and Planck combined with LIM with $\sim 10^8$ spectrometer-hours.}
\begin{center}
\noindent\scriptsize
\begin{minipage}[t]{0.45\textwidth}
\centering
\renewcommand{\arraystretch}{1.2}
\begin{tabularx}{.85\textwidth}{*{4}{C}} 
\multicolumn{4}{c}{{\boldmath $\Lambda {\rm CDM}+ N_{\rm eff}$}}  
\vspace{.04in} \\  
\hline 
Parameters & Planck & +Euclid & +LIM  \vspace{.04in} \\ 
\hline 
{\boldmath$\ln (10^{10}A_s)$} & 0.018 & 0.0035 & 0.0029 \\
{\boldmath$n_s$}& 0.0085 & 0.0033 & 0.0016\\
{\boldmath$h$}&  0.014 & 0.0056 & 0.0019\\
{\boldmath$\Omega_b$}& 0.0017 & 0.00065 & 0.00030 \\
{\boldmath$\Omega_c$}& 0.0086 & 0.0031 & 0.0014\\
{\boldmath$N_{\rm eff}$} & 0.187 & 0.073 & 0.023\\ \\
\end{tabularx} 
\end{minipage}
\begin{minipage}[t]{0.45\textwidth}
\centering
\renewcommand{\arraystretch}{1.2}
\begin{tabularx}{.85\textwidth}{*{4}{C}} 
\multicolumn{4}{c}{{\boldmath $\Lambda {\rm CDM}+ M_\nu$}} 
\vspace{.04in} \\
\hline  
Parameters & Planck & +Euclid & +LIM  \vspace{.04in} \\ 
\hline 
{\boldmath$\ln (10^{10}A_s)$}& 0.015 & 0.012 & 0.0055\\
{\boldmath$n_s$}& 0.0044 & 0.0029 &  0.0014\\
{\boldmath$h$}&  0.011 & 0.0024  & 0.0013 \\
{\boldmath$\Omega_b$}& 0.0014 & 0.00035 & 0.00018 \\
{\boldmath$\Omega_c$}& 0.011 & 0.0024 & 0.0011\\
{\boldmath$M_\nu$} [eV] & 0.083 & 0.031 &  0.013\\ \\
\end{tabularx} 
\end{minipage}
\vspace{.1in}

\begin{minipage}[t]{0.45\textwidth}
\centering
\renewcommand{\arraystretch}{1.2}
\begin{tabularx}{.85\textwidth}{*{4}{C}} 
\multicolumn{4}{c}{{\boldmath $\Lambda {\rm CDM} + N_{\rm eff}+M_\nu$}}  
\vspace{.04in} \\ 
\hline  
Parameters &  Planck & +Euclid & +LIM \vspace{.04in} \\ 
\hline 
{\boldmath$\ln (10^{10}A_s)$}& 0.018 &  0.013 & 0.0063\\
{\boldmath$n_s$}& 0.0088 & 0.0037 & 0.0016 \\
{\boldmath$h$}& 0.018 & 0.0056 & 0.0021 \\
{\boldmath$\Omega_b$}& 0.0022 & 0.00067 & 0.00033\\
{\boldmath$\Omega_c$}& 0.012 & 0.0031 & 0.0014 \\
{\boldmath$M_\nu$}\ [eV] & 0.094 & 0.035 & 0.015 \\
{\boldmath$N_{\rm eff}$}& 0.192 & 0.081 & 0.030\\ \\
\end{tabularx} 
\end{minipage}
\begin{minipage}[t]{0.45\textwidth}
\centering
\renewcommand{\arraystretch}{1.2}
\begin{tabularx}{.85\textwidth}{*{4}{C}} 
\multicolumn{4}{c}{{\boldmath $\Lambda {\rm CDM} + N_{\rm eff}+Y_{\rm He}$}} \vspace{.04in} \\ 
\hline 
Parameters &  Planck & +Euclid & +LIM \vspace{.04in} \\ 
\hline 
{\boldmath$\ln (10^{10}A_s)$}& 0.019 & 0.0040 & 0.0032\\
{\boldmath$n_s$}& 0.0086 & 0.0037 & 0.0016  \\
{\boldmath$h$}& 0.018 &  0.0059 &  0.0019 \\
{\boldmath$\Omega_b$}& 0.0027 & 0.00089 & 0.00038 \\
{\boldmath$\Omega_c$}& 0.0096 & 0.0031 &  0.0014 \\
{\boldmath$Y_{\rm He}$} & 0.018 & 0.012 & 0.0079\\
{\boldmath$N_{\rm eff}$} & 0.30 & 0.10 &  0.029\\ \\
\end{tabularx}
\end{minipage}
\vspace{.1in}

\begin{minipage}[t]{0.45\textwidth}
\centering
\renewcommand{\arraystretch}{1.2}
\begin{tabularx}{.85\textwidth}{*{4}{C}} 
\multicolumn{4}{c}{{\boldmath $\Lambda {\rm CDM} + M_\nu+w_0$}} 
\vspace{.04in} \\ 
\hline  
Parameters &  Planck & +Euclid & +LIM \vspace{.04in} \\ 
\hline 
{\boldmath$\ln (10^{10}A_s)$}& 0.016 & 0.014 &0.0058 \\
{\boldmath$n_s$}& 0.0055 & 0.0034 & 0.0014 \\
{\boldmath$h$}& 0.091 & 0.0058& 0.0014 \\
{\boldmath$\Omega_b$}& 0.0054 &  0.00098 & 0.00013\\
{\boldmath$\Omega_c$}& 0.023 & 0.0034 & 0.0010 \\
{\boldmath$w_0$}& 0.28 & 0.018  & 0.0051 \\
{\boldmath$M_\nu$}\ [eV] & 0.10 &0.043 & 0.017 \\ 
& & &  \\ \\
\end{tabularx} 
\end{minipage}
\begin{minipage}[t]{0.45\textwidth}
\centering
\renewcommand{\arraystretch}{1.2}
\begin{tabularx}{.85\textwidth}{*{4}{C}} 
\multicolumn{4}{c}{{\boldmath $\Lambda {\rm CDM} + M_\nu+w_0+w_a$}} 
\vspace{.04in} \\ 
\hline 
Parameters &  Planck & +Euclid & +LIM \vspace{.04in} \\ 
\hline 
{\boldmath$\ln (10^{10}A_s)$}&0.016 & 0.013 & 0.0057\\
{\boldmath$n_s$}& 0.0045& 0.0029& 0.0015 \\
{\boldmath$h$}& 0.097&  0.0054 & 0.0019 \\
{\boldmath$\Omega_b$}& 0.0044 & 0.00026 &  0.00014 \\
{\boldmath$\Omega_c$}& 0.037 &  0.0026&  0.0014 \\
{\boldmath$w_0$}& 0.42 & 0.027  & 0.0098\\
{\boldmath$w_a$}&  1.0  &  0.096 &  0.041 \\
{\boldmath$M_\nu$}\ [eV] & 0.11  &  0.045 & 0.024 \\ \\
\end{tabularx}
\end{minipage}
\vspace{.1in}
 \end{center}
 \label{tab:1sig}
\end{table*}

\begin{figure*}[!htbp]
    \centering
    \includegraphics[width=.65\textwidth]{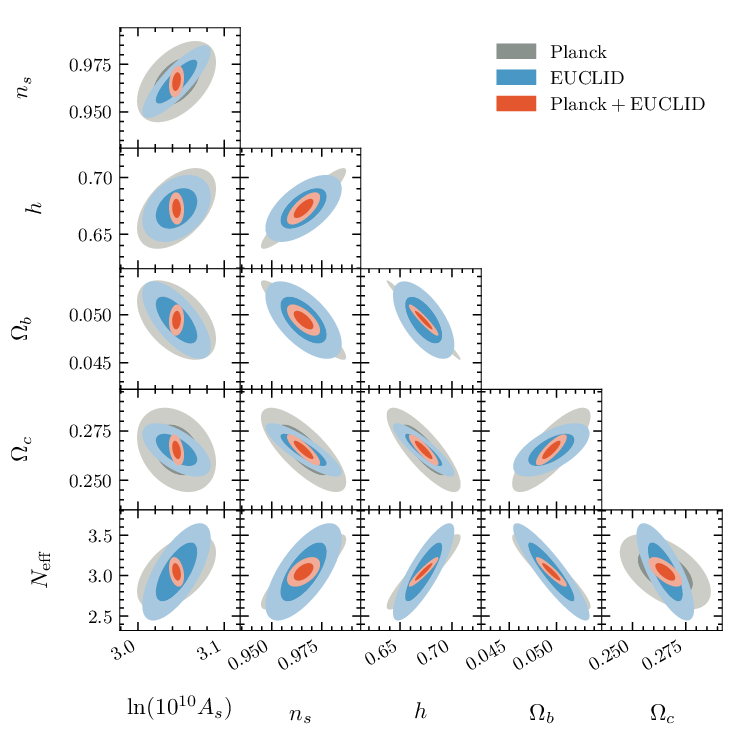}\vspace{-.15in}
    \includegraphics[width=.65\textwidth]{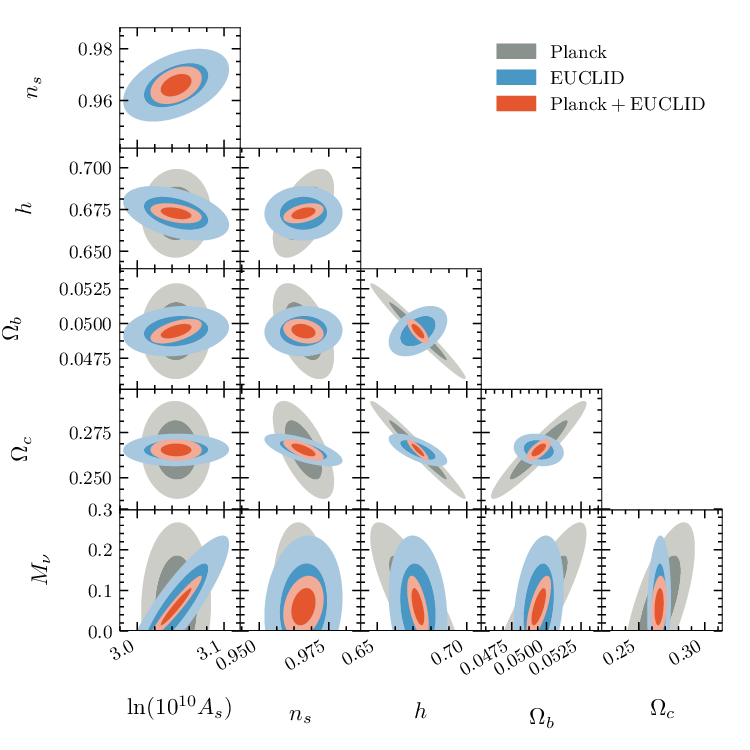}\vspace{-.2in}
    \caption{2D marginalized constraints on cosmological parameters for $\Lambda{\rm CDM}+N_{\rm eff}$ (top) and $\Lambda{\rm CDM}+M_\nu$ model (bottom) from Planck (gray) and Euclid (blue) data alone and combined (red).}
    \label{fig:EUCLID_triangle}
\end{figure*}

\clearpage
\newpage
\bibliography{LIM_neutrinos}

\end{document}